\documentclass[a4paper,12pt]{article}
\pdfoutput=1 

\usepackage{jheppub,xcolor}
\usepackage{natbib}
\usepackage[T1]{fontenc} 
\usepackage{subfigure}
\usepackage{tabularx}
\usepackage{comment}
\usepackage{appendix}
\title{Confined phases at finite density in the Hardwall model}
\author{Akash Singh, K. P. Yogendran}
\affiliation{Department of Physical Sciences,\\
IISER Mohali\\Sector 81, Knowledge City\\Punjab, 140306, India}

\emailAdd{akashsingh@iisermohali.ac.in, yogendran@iisermohali.ac.in}

\abstract{AdS/QCD models are being extensively studied because they seem to offer an entirely tractable and radically different approach to the phases of QCD whose region of validity is precisely the region of strong interactions. In this context, a missing ingredient has been the absence of a bulk description dual to a {\em confined} hadronic phase at finite baryon density. In this work, we discuss a class of backgrounds that fill this lacuna.  Applying physically motivated boundary conditions, we obtain phase diagrams which compare reasonably with approaches based on the PNJL model. We also show that such backgrounds can play a similar role in geometries with a dilaton field.
}

\makeatletter
\gdef\@fpheader{}
\makeatother

\begin{document} 
\maketitle
\flushbottom

\section{Introduction}

Condensed QCD matter, especially at intermediate densities and temperatures, presents a complex phase diagram \cite{Sedrakian:2018ydt, Jaiswal:2020hvk,Gross:2022hyw, Aarts:2023vsf,kumar2024theoretical} 
with diverse properties that depend sensitively on the quark content, chiral symmetry, axial anomaly, strong coupling, and topological charge \cite{Cui:2021bqf, Pisarski:2024esv}. Additionally, calculational difficulties due to strong interactions mean that the thermodynamic and transport properties of these phases of matter have not yet been determined with sufficient confidence and precision. This is especially relevant in confronting theoretical predictions with the high precision multimessenger pulsar data, which must be explained in terms of these properties \cite{Blaschke:2018mqw}. 

While various phenomenological approaches such as chiral perturbation theory \cite{Weinberg:1978kz,Gasser:1983yg,Ananthanarayan:2023gzw}, PNJL model \cite{Meisinger:1995ih,Fukushima:2003fw}, Hadron Resonance Gas (HRG) model \cite{Hagedorn:1965st,Ratti2021} etc. exist, these are not definitive in the region of interest because of multiple reasons: the validity limits of the approximations being made are not transparent, the use of 3-momentum cutoffs can lead to equations of state that are acausal $c_s^2>1$ \cite{Baym:2017whm,Jeong:2019lhv}, and thermodynamic consistency (such as positivity of transport coefficients) is not guaranteed especially in applications which depends on the details of transport and relaxation properties \cite{Lindblom:2018rfr}. One reason being that at these intermediate densities, QCD possesses multiple dynamical scales making it difficult to study.

Lattice Gauge theory, being a first principles approach, can compute the equations of state exactly, but convergence issues related to the sign problem have presented a significant barrier. Recent progress summarized in \cite{Nagata:2021ugx,ratti2023equation,Aarts:2023vsf} shows that much work remains before these can be confronted with experimental data at intermediate densities and temperatures. 

Given this status, it is important to explore alternative models that can aid in developing a causal and thermodynamically consistent understanding of these phases. It is in this context that the AdS/QCD holographic approach can be of significance. This approach is surprisingly consistent with the QCD phenomenology in the regions (of temperature, density, and couplings) where it has been compared \cite{Kim:2012ey,Ammon:2015wua}, and provides a technically simpler framework. Mesons \cite{Erdmenger:2007cm,MartinContreras:2021yfz,Afonin:2021cwo} and baryons \cite{Hong:2007ay,Fang:2016uer,Bigazzi:2018cpg,Hashimoto:2019wmg,Kovensky:2021ddl,Jarvinen:2022gcc} have been extensively studied in holographic AdS/QCD models, together with other exotic boundary field theory matter states \cite{Elliot-Ripley:2016uwb,Hartnoll:2016apf}. Symmetry energy \cite{Park:2011zp,Bartolini:2022gdf}, chiral phase transition \cite{Chelabi:2015gpc,Li:2016smq}, transport coefficients \cite{Hubeny:2011hd,Casalderrey-Solana:2011dxg,Hoyos:2021njg} are some contributions of the vast number of studies in this subject.

Also, significant amount of work has been done to obtain the EOS of dense QCD matter at intermediate densities and temperatures using holography. Extensive explorations \cite{BallonBayona:2007vp,Karch:2007br,Megias:2010ku,DeWolfe:2010he,Kim:2014pva,Hoyos:2016zke,Knaute:2017opk,Jokela:2018ers,Annala:2019puf,Mamani:2020pks,BitaghsirFadafan:2020otb, Ghoroku:2021fos,Hippert:2023bel,Braga:2024nnj}, both foundational and phenomenological, have shown that these methods yield a radically new description of strongly coupled matter that is nevertheless thermodynamically consistent as well as consistent with various causality and unitarity constraints that are expected to be applicable. The foundational models, referred to as the top-down approach, involve 10-dimensional geometries derived from string theory first principles. Notable examples of this approach include the D3/D7 setup \cite{Karch:2002sh,    BitaghsirFadafan:2020otb} and the Witten-Sakai-Sugimoto models \cite{Kovensky:2021kzl,Rebhan:2014rxa}, in which probe flavor D7-branes (D8-branes) are studied in the spacetime background produced by D3-branes (D4-branes).

Among the holographic models, two simple phenomenology driven approaches are the hardwall and the softwall models \cite{Erlich:2005qh,Karch:2006pv}. In the former, AdS conformal invariance is explicitly broken by introducing a cutoff along the AdS radial direction. In the latter, conformal invariance is broken by a nontrivial scalar field profile, which has been suggested to be the dual of a running coupling in the field theory \cite{Kehagias:1999tr}. An important line of exploration in holographic QCD is the VQCD models \cite{Jarvinen:2021jbd}. These models, inspired by the softwall models of IHQCD \cite{Gursoy:2007cb,Gursoy:2007er}, incorporate flavor degrees of freedom by accounting for backreaction with tachyon brane actions \cite{Bigazzi:2005md,Casero:2007ae} within the Veneziano limit ($N_c \to \infty$ and $N_f \to \infty$) while keeping the ratio $\frac{N_f}{N_c}$ fixed. 

Much of the recent studies consider only AdS black hole backgrounds and thus do not consider the possibility of a {\em confined} phase at finite baryon density. In this work, we remedy this situation by discussing a family of background geometries which we call the charged AdS geometry (CAdS). These have nonzero electric fields threading the AdS radial direction, but do not have a black hole. Thus, the entropy will not contain an $\mathcal{O}(N_c ^2)$ term (and the Polyakov loop expectation value will vanish \cite{Colangelo:2010pe}) indicating a confined phase that describes the system at low temperatures and densities. Similar backgrounds were first discussed in \cite{Lee:2009bya,Sachan:2013zza} but these have not received as much attention as they should in this context. In our work, we utilise these backgrounds to build a complete phase diagram in the $\m -T$ plane. We also explain how to use physically motivated boundary conditions at the hardwall cutoff.

The action and the procedure used to fix the parameters in the action are described in the next section. We then present the various backgrounds that we consider including the new CAdS background and its properties in section \ref{solutions}. We extend the confinement/deconfinement phase transition to finite chemical potential in section \ref{phasetransition}.
In section \ref{PIR}, we will also show how the hardwall approach can be significantly advanced by using phenomenologically motivated IR boundary conditions. These lead to sensible phase diagrams showing much promise. We follow this up by showing that CAdS backgrounds can also be constructed in the nontrivial dilaton geometries in section \ref{dilaton}. Finally, we summarize and discuss future directions that merit exploration in arriving at robust and predictive constructions to study condensed QCD matter at finite density.

\section {Hardwall model}\label{HW}
The hardwall approach requires us identify the lowest Free energy solutions of the equations of motion obtained from an action describing a Quantum Gravity theory on an asymptotically AdS background constrained by specified boundary conditions. The power of this approach derives from its use of extremely simple actions with few parameters - we will start with the action
\be\label{action1}
S=\frac{1}{2\k^2}\int_0 ^{z_0}dz\, d^4x\, \sqrt{g}\left(R-2\L\right) 
- \frac{1}{g_5^2}\int_0 ^{z_0} dz\, d^4x\, \sqrt{g} \frac{F^2}{4 }.
\ee
where the integration range along the radial AdS direction is restricted by the IR cutoff $z_0$ and,  
\be\label{kg} 
\k^2=\frac{4\p^2 L^3}{N_c^2}; \qquad g_5^2=\frac{24 \p^2 L}{\q N_f N_c }
\ee 
The action for the $U(1)_B$ gauge field can be regarded, in string theoretic terms, as arising from the Yang-Mills approximation to the DBI action of flavor D7-branes wrapping an $S^3$ inside a ten dimensional $AdS_5\times S^5$ spacetime. In this case, the dimensionful 5D gauge coupling $\frac{1}{g_5 ^2}=N_f\t_7L^3 2\pi ^2 (2\p\a')^2=\frac{ N_c N_f}{4\pi^2 L}$. More generally, the D7-brane can be wrapping various compact cycles (such as in the Klebanov-Strassler setting) and a dilaton background could also be present. Therefore, we can introduce a parameter $\q$ to remind us of these possibilities and a large value for $\q$ indicates the significance of the effects from the compactification manifold and/or dilaton. 
We will set $\q=6\sqrt{\frac{2N_c}{N_f}} , 13.5$, respectively where the former value is determined by equating the free energy at high densities with the Stefan-Boltzmann value \cite{Hoyos:2016zke}. Corresponding to these values of $\q$ , the parameter defined below equals $\z=0.77, 1$ respectively and the latter value  is chosen for representation purposes only. 

The cosmological constant in the Einstein-Hilbert part of the action is set to be $2\L=-\frac{12}{L^2}$, so that the asymptotic solution is $AdS_5$ with the AdS length scale set by $L$.  

Motivated by translational and rotation symmetry of the phases we are interested in, the backgrounds we consider involve a diagonal metric ansatz,
 \be 
 ds^2=-\frac{g(z)}{h(z)}L^2 dt^2+ \frac{L^2}{z^2}d\Vec{x}^2+\frac{L^2}{g(z)} dz^2
 \ee
which implies that the (conformal) boundary is at $z=0$. 
The gauge field ansatz is $A_0=\phi(z)$, $A_{i}=0$ where $i=1,2,3$ with gauge condition $A_z=0$. The equations of motion with these ans\"atze are: 
\bea
&&g'(z)-g(z) \left(\frac{4}{z}+\frac{h'(z)}{2 h(z)}\right)+4 z-2\z z \left(\frac{ h(z) \phi '(z)^2}{4} \right)=0\label{EOMg} \\
&&\frac{h'(z)}{h(z)}-\frac{4}{z}=0\label{EOMh} \\
&&\phi ''(z)-\phi '(z) \left(\frac{3}{z}-\frac{h'(z)}{2 h(z)}\right)=0\label{EOMph}
\eea
where the parameter $\z=\frac{2 \k^2}{3 g_5^2}=\frac{\q N_f}{9 N_c}L^2$ controls the effect of the gauge field on the background (large $N_c$ being the probe approximation).  We will show in section \ref{PIR} that there exists a small range of $0.75\lesssim \z\lesssim 1.25$ for which the phase diagram has sensible QCD interpretation. 

The work of \cite{Herzog:2006ra} explains how breaking scale invariance in 5 dimensions leads to a deconfinement transition at zero density as we vary the temperature. The transition temperature can be translated to physical units if the parameters appearing in the gravity action are fixed in terms of phenomenology. This can be done by following the work of \cite{Erlich:2005qh}. Since the parameters are being fixed by few particle quantum states, the deconfinement temperature is a {\em prediction} upto thermal loop corrections (but there are other more important but well known limitations as mentioned in the discussion section). We will now explain how these ideas can be extended to the study of the finite density phases of these hardwall models. 

We start by fixing the parameters of the Lagrangian in terms of physical quantities. The IR cutoff $z_0$ is fixed by computing the {\it glueball} mass in the {\it absence} of the flavor degrees of freedom. That is, we assume that the bulk metric is 
\be
ds^2=\frac{L^2}{z^2}(dx^2+dz^2) \quad 0<z\leq z_0
\ee
and compute the spectrum of glueball excitations \cite{Rinaldi:2017wdn}. This is equivalent to ignoring the effects of the quarks on the glueball masses. 

On the other hand, we can regard the gauge fields as describing the dynamics of the open string modes on the flavor branes in a string theoretic perspective. These are, then,  governed by the open string metric $g_o=P[G]$ which depends on the shape of the branes via the pullback. Thus, in the general situation, $g_0$ can be expected to depend on $N_f$ and the quark mass and the chiral condensate. Hence for these we may take the metric to be 
\be
 ds^2=\frac{L^2}{z^2}\left(-\frac{g(z)}{z^2}dt^2+ d\Vec{x}^2+\frac{z^2}{g(z)} dz^2\right)
\ee
which permits an IR boundary condition $g_0=g(z_0)$ different from that of the closed string modes above. We will set $g(z)=z^2(1-c z^4)$ because this solves the bulk equations of motion (if $c>0$, the geometry behind the cutoff will have a blackhole though).

We will fix the parameter $g_0$ by computing the vector meson mass \cite{Erlich:2005qh} using an $SU(N_f)_V$ (isospin) gauge field. We consider turning on the 3-component  $a^{(3)} _\m$ and search for a normalizable  plane wave fluctuation:
\be
\frac{1}{\sqrt{g}}\del_z(\sqrt{g}g^{zz}g^{\n\n} f_{z\n})-g^{\m\m}g^{\n\n}f_{\m\n}(k,z)=0.
\ee
This solution must have no nodes in the radial direction which is uniquely determined given the boundary conditions 
\be
a_\n(k,0)=\del_z a_\n(k,z_0)=0 \qquad 
\ee
The latter Neumann boundary condition is consistent with the residual gauge transformations in the gauge $a_z=0.$ Because of the residual gauge symmetry, the solution must take the form 
\be
a_\nu=(k_\n k_\m-\h_{\m\n}k^2)\e^\m_{k}(z) e^{ikx}
\ee
where $\e_k ^\m(z),$ which depends only on $k^2$ by Lorentz invariance along the boundary directions, satisfies
\be
z\del_z(\frac{g}{z^3}\del_z V(z))+k^2 V(z)=0 \label{rho-mass}
\ee
where $V_\n(z)=k^2\e_\n-k_\n k^\s\e_\s$ for some component $\n.$
Since the $\rho-$meson mass is 776 MeV this means $k^2 z_0^2=m_\rho^2 z_0^2=q^2 $, here $q$ is the eigenvalue of the differential operator. Equating $z_0 ^{-1}=290$ MeV, we get $g_0\simeq6.$ Since $g_0>1$, there is no horizon even behind the cutoff. For $z_0^{-1}=250$ MeV, $g_0\simeq27.$ As we will see later, these parameters do not change the phase diagram qualitatively but will affect the phase boundaries. 

We will also consider the AdS black hole background and in this case, setting $z_0=z_H$ will imply $g_0=0$ which is appropriate to a situation with {\em chiral symmetry being restored}. 

\section{Finite $\mu_B$ backgrounds}\label{solutions}
To begin with, we list the known possibilities at finite {\em quark} chemical potential $\m_Q=\mu =\frac{\mu_B}{N_c} $ which is the non-normalizable mode of the gauge field $\f.$  (This is consistent with baryons being represented as wrapped branes and the normalization of the DBI action). The equations of motion \eqref{EOMg}, \eqref{EOMh}, \eqref{EOMph} 
obtained from the Lagrangian \eqref{action1} can be solved analytically. Two solutions have been studied in the literature, namely,
\begin{itemize}
    \item Thermal AdS (thAdS)
        \be   
        ds^2=\frac{1}{z^2}(-dt^2+d\Vec{x}^2+dz^2)
        \ee
        \be 
        \phi(z)=\m
        \ee 
\end{itemize}
with zero charge density and pressure
\be
p=\frac{N_c ^2}{8\pi^2}\biggl(\frac{z_0 ^2 + g(z_0)}{z_0 ^6}\biggl)
\ee
which is computed using the Holographic renormalization method as explained in Appendix \ref{holoren}. The second background is the
\begin{itemize}
    \item Charged black hole (CBH)
        \be  \label{CBH} 
        ds^2=\frac{1}{z^2}\left(-f dt^2+d\Vec{x}^2+\frac{dz^2}{f}\right)
        \ee 
        where $f=1-c z^4 + \frac{\z \mu ^2}{z_H^4 L^2}z^6$ 
        \be 
        \phi(z)=\m(1-\frac{z^2
}{z_H ^2})        \ee
where $z_H$ is the black hole horizon. The parameter $c$ is fixed by the condition that $z_H$ is the smallest root of $f$ for each $\mu.$
\end{itemize} 
The Gibbons-Hawking argument determines the periodicity of the Euclidean time coordinate (and hence the temperature) in the black hole geometry 
\be
T_H=\frac{4c z_H^3-6\z \m^2 z_H}{4\pi}=\frac{1}{\pi z_H}\Big|1- \frac{\z \m^2 z_H^2}{2L^2}\Big|
\ee
while the temperature of the thAdS can be set to any value following the same argument provided in \cite{Herzog:2006ra} or in \cite{Singh:2022obu}. The temporal component of the gauge field $\phi(\e)$ approaches the same constant $\m$ near the boundary for both geometries. In this case, 
since the electric field is sourced by the black hole, the baryon number density can be interpreted to be in the deconfined phase. This is because quarks are described in the bulk as open strings that end on the boundary. If there is a black hole, the other end of these open strings can end on the horizon with a finite energy (after renormalization). Calculating the pressure (see Appendix \ref{holoren}) we get, 
\be
p=\frac{N_c^2}{8\pi^2}\biggl(\frac{1}{z_H ^4}+\frac{\z\mu^2}{z_H ^2 L^2}\biggl).
\ee
The zero temperature limit of this pressure $p=\frac{\q^2 N_f^2}{864\pi^2}\mu^4$ is proportional to $N_f ^2$ and vanishes as $\mu\to 0.$ Remarkably, there are no `gluonic' or mesonic terms in the pressure (which will be proportional to $N_c ^2$ and $N_c N_f$ respectively).

\subsection{Charged AdS}

However, at small densities, we can expect that the dual field theory has a non-degenerate ground state which can be expected to be described by a horizonless geometry in holography. To identify such a background,  we observe that the function $f$ of the preceding section does not vanish when $\frac{\z^2 \m ^{4}}{L^4 z_H^8 c ^3}>\frac{4}{27}.$
These ``overcharged" geometries thus expose the IR singularity which occurs at 
$z\to\infty$ and are usually disregarded. However, in our present considerations, the hardwall IR cutoff $z_0$ will screen the singularity and hence these backgrounds maybe allowed. A similar model was considered in \cite{Lee:2009bya}, however, with $c=0$.

Thus, we obtain another class of backgrounds which we call Charged AdS.
\begin{itemize}
\item Charged AdS (CAdS)
\be \label{CAdS} 
ds^2=\frac{1}{z^2}(-fdt^2+d\Vec{x}^2+\frac{dz^2}{f})
\ee 
where $f=(1-c z^4 +\frac{ \z}{L^2} Q^2 z^6)$ with $\frac{\z^2 Q ^{4}}{L^4c ^3}>\frac{4}{27}.$
\be 
\phi(z)=\m- Q z^2
\ee
\end{itemize}
We assume that since these geometries do not have a horizon, they will also not host {\em free} quark degrees of freedom. Hence, these CAdS geometries must be interpreted as describing a finite density confined phase.

In CAdS geometries, we need a boundary condition at the IR cutoff
that will result in an equation of state. A simple choice of boundary conditions on the gauge field is to set $\f(z_0)=0$ which leads to an equation of state $Q=\frac{\mu}{z_0 ^2}$ where $Q$ has dimensions of number density. On the other hand, if there are sources for the gauge field  present behind the cutoff surface, other boundary conditions maybe more natural as we will discuss in Section \ref{PIR}.

For the charged AdS geometry, $c=\frac{1}{z_0 ^4}-\frac{g(z_0)}{z_0 ^6}+\frac{\z Q^2z_0 ^2}{L^2},$ and so we obtain
\be\label{pcads}
p=\frac{N_c ^2}{8\pi^2}\biggl(\frac{z_0 ^2 + g(z_0)}{z_0 ^6}+\frac{\z Q^2z_0 ^2}{L^2}\biggl).
\ee
The $z_0 ^2$ contributions in the pressure can be attributed to the glueballs since it will be present even in the absence of fundamental matter. It does not depend on density (and in AdS backgrounds it does not depend on temperature either) and can be subtracted away as is often done in the literature.
However, $g_0$ depends on the fundamental degrees of freedom and will include effects of the chiral condensate and possibly mesons as well. These different contributions maybe disentangled in a more nuanced model such as the 10-d hardwall model where the D7-branes will also contribute to $g_0$. In this sense, our model is similar to the those presented in \cite{Bartolini:2022rkl}, \cite{Singh:2022obu} where independent hardwall cutoffs were considered for the flavor and color degrees of freedom. However, in this work we have equated the two cutoffs and hence the numerical value of $g_0$ obtained by the fit does not make  factors of $\frac{N_f}{N_c}$ explicit. We will therefore retain the $g_0$ term assuming it to capture the contribution of fundamental matter.

We note that in the above geometry, because of the IR terms, the pressure does not vanish at zero density (and temperature). 
If we use the Euler relation $\e=\mu\frac{\del p}{\del \mu}-p,$
to determine the energy density, we find that 
\be
\e=\frac{N_c ^2}{8\pi^2}\biggl(\frac{2\z\mu Q Q'z_0 ^2}{L^2}-\frac{z_0 ^2 + g(z_0)}{z_0 ^6}-\frac{\z Q^2z_0 ^2}{L^2}\biggl)
\ee 
which becomes negative at zero density (since $g_0>0$). This is an unavoidable feature of the hardwall models and implies that these do not capture the physics near $\mu, T=0.$  

The second term in the pressure 
\be
\frac{N_c ^2}{8\pi^2 L^2} \z Q^2z_0 ^2\sim N_c N_f Q^2
\ee
arises from the baryon density $\rho\sim Q Q'$ and hence depends on $Q(\mu,T)$. In a more complete holographic scenario, we can expect that this term arises due to dynamics of solitons {\em localized} in the IR part of the flavor branes such as D5-branes in $AdS_5$ \cite{Brandhuber:1998xy},\cite{Imamura:1998hf} and wrapped D4-branes in the $D4/D8$ WSS models \cite{Kovensky:2021ddl}. In our hardwall approach, these are assumed to be localized in the region beyond the IR cutoff $z_0.$
Thus, we may choose $Q(\mu,T)$ so that this term in the pressure models the baryon pressure computed by other methods. We will discuss this in section \ref{PIR}.

It maybe remarked that these geometries can be regarded as a softwall model as well. In the usual softwall models, an IR region of large {\em gravitational} curvature and string coupling plays the role of a wall for the gravity degrees of freedom. In the present case, a softwall arises because of the growth of the electric field due to the baryon density. Then, the curvature of the open string metric becomes large and therefore will lead to a wall for the fundamental degrees of freedom. 

\section{Deconfinement at finite density}\label{phasetransition}

We now show that the phase transition between confined and deconfined phases of \cite{Herzog:2006ra} can be extended to finite chemical potential. 
The CAdS has always the lower free energy than thAdS, and we will not consider it further. The difference between pressures of the confined CAdS background and the charged black hole background is,
\be 
p_{CAdS}-p_{CBH}=\frac{L^3}{2\k^2}\left(\frac{g_0}{z_0^6}+\frac{1}{z_0 ^4}- \frac{1}{z_H^4}+ \z  \left( \frac{Q^2z_0^2}{L^2}- \frac{\mu^2}{ z_H^2}\right) \right).
\ee  
At zero temperature $\frac{\z \m^2 z_H^2}{2L^2}=1$ and this difference is negative when 
\be
(\mu z_0)^2>\frac{2 L^2}{3\z}\left(1+\sqrt{4+3\frac{g_0}{z_0^2}}\right). \label{mucrit}
\ee
In this inequality which determines the deconfinement transition, we can interpret $z_0$ as the mass of the glueball (or equivalently $\L_{QCD}$) while $g_0$ is linearly related to the meson mass (see \eqref{rho-mass}).
\be
\left(\frac{\mu}{\a M_g}\right)^2>\frac{6 N_c}{\q N_f}\left(1+\sqrt{4+3\b \left(\frac{M_\rho}{\a M_g}-1.766\right)^4}\right)
\label{muCRIT_QCD}
\ee
 where $\a \simeq 0.26,0.195$ for Dirichlet and Neumann boundary conditions, respectively, and $\b\simeq 8.37$ are numerical constants. It will be very interesting to check this against the experimental data for suitable choices of $\a$ and $\b$. 

We now recall that while the pressure in the CBH geometry varies with temperature via its dependence on the horizon position, it is independent of temperature in the CAdS geometry. Even so, comparing the pressures of these two geometries will produce a temperature dependent transition. 
For both backgrounds, we can compute finite temperature corrections from {\em bulk} thermal {\em loops}, but these will be $N_c$ suppressed. 

\begin{figure}[h]
    \centering
    \subfigure[]{\includegraphics[width=0.45\textwidth]{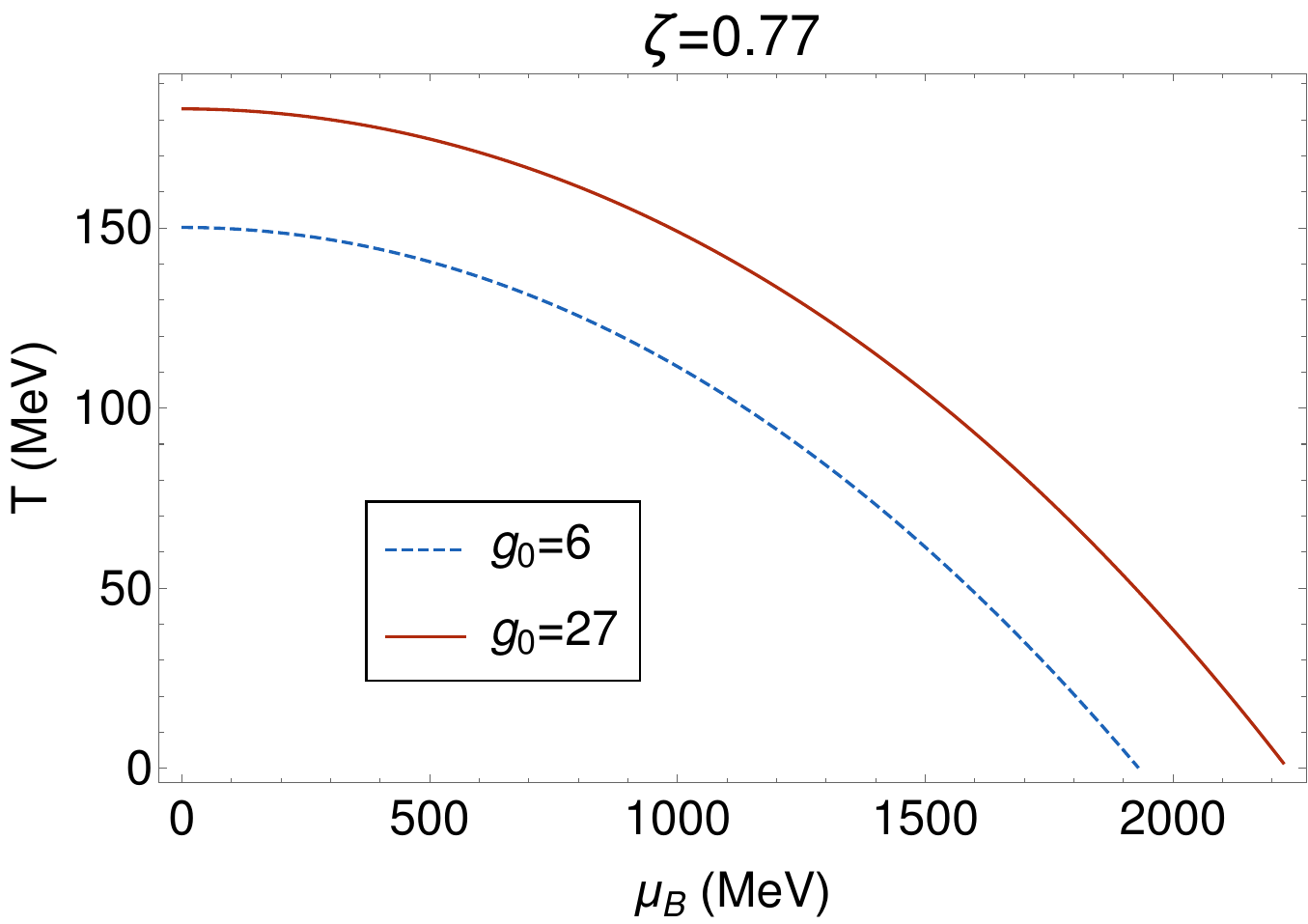}}
    \subfigure[]{\includegraphics[width=0.45\textwidth]{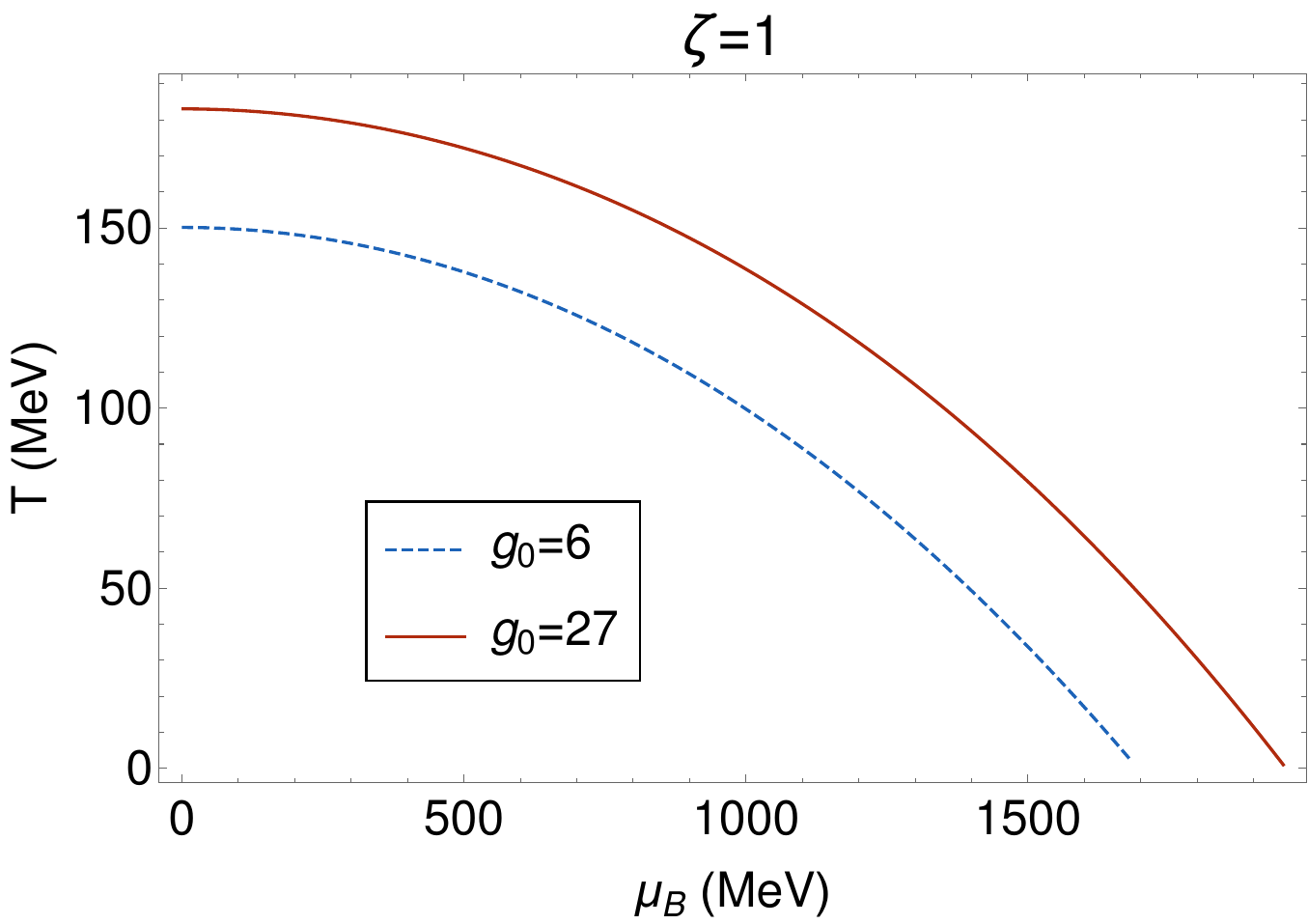}}
    \caption{Phase Diagram for different $\z.$}
    \label{fig:pdCBHCADSzeta}
\end{figure}

Translating into physical units as explained earlier, we get the phase diagram as shown in figure \ref{fig:pdCBHCADSzeta} for different values of $g_0$ at fixed $\z$.
A greater $g_0$ leads to an increased critical temperature and chemical potential. These diagrams create the impression that $g_0$ functions merely as a scale and that all the curves are identical.

\begin{figure}[h]
    \centering
    \subfigure[]{\includegraphics[width=0.45\textwidth]{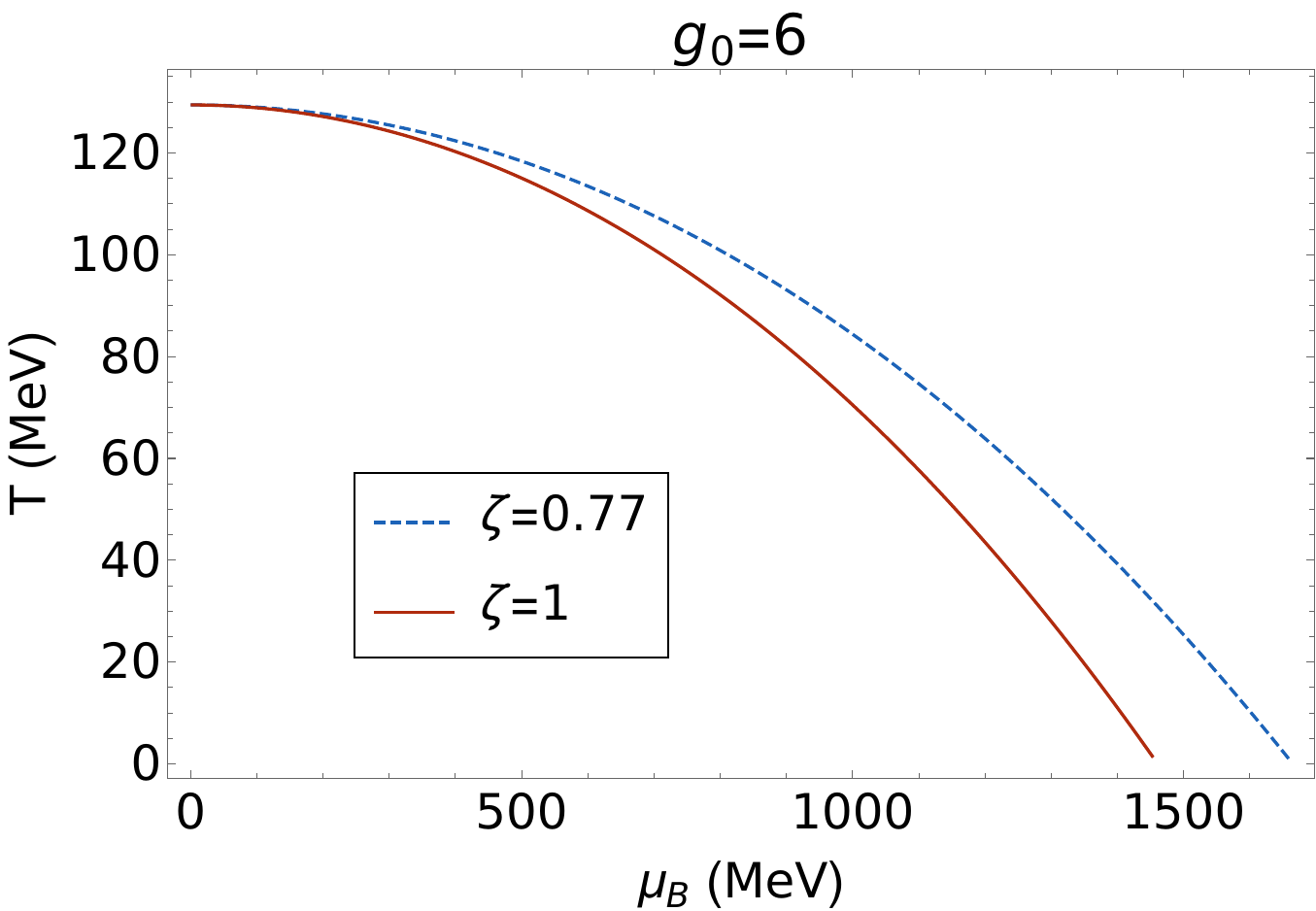}}
    \subfigure[] {\includegraphics[width=0.45\textwidth]{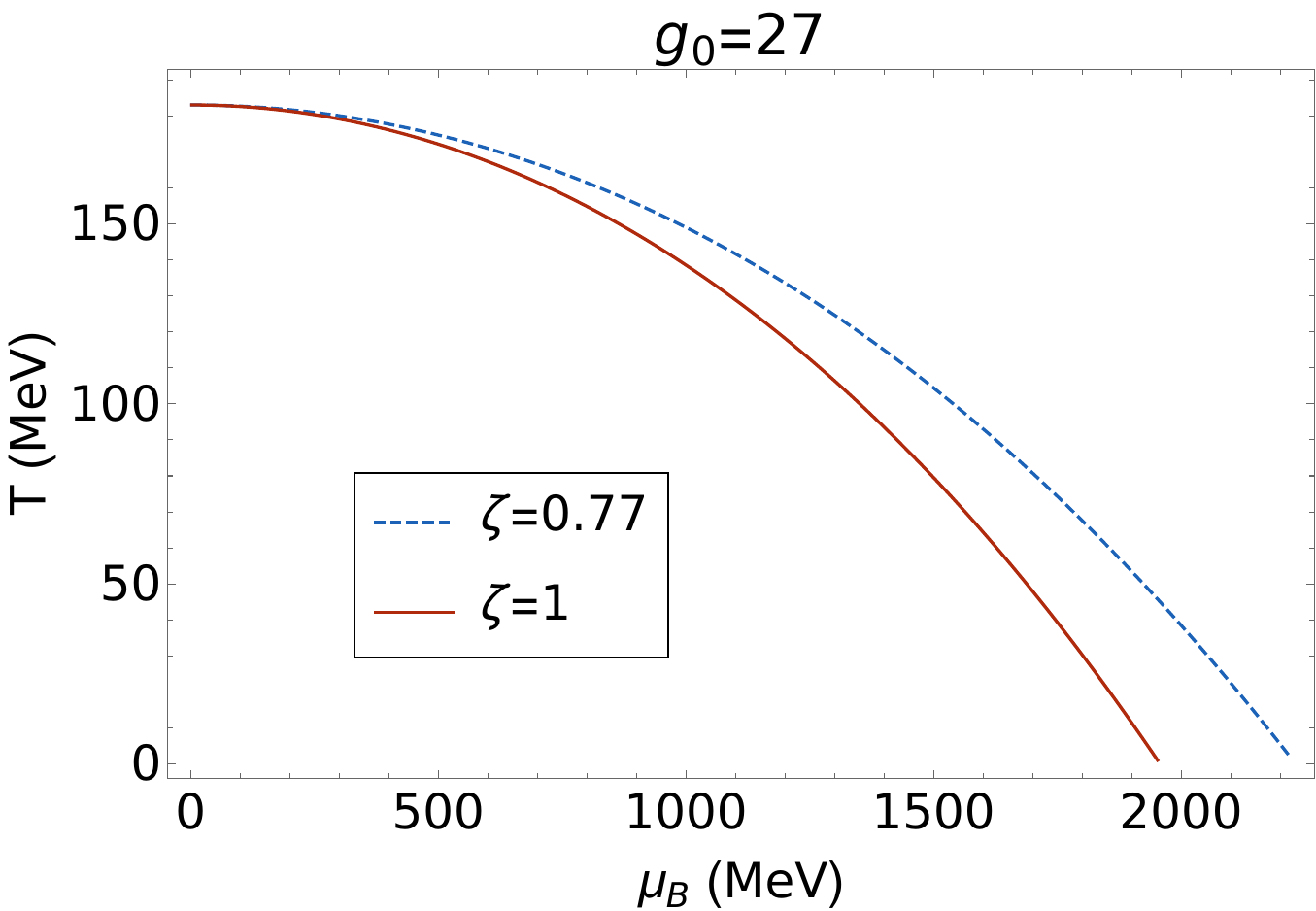}}
    \caption{Phase Diagram for different $g_0$ }
    \label{fig:pdCBHCADSg0}
\end{figure}

Figure \ref{fig:pdCBHCADSg0} shows the behavior as we vary $\z$ while keeping $g_0$ fixed. 
It is clear that a larger $\z$ produces lower critical chemical potential because of the inverse scaling \eqref{mucrit}. However, at zero chemical potential, the density vanishes and we recover the phase transition of \cite{Herzog:2006ra}.

\subsection{Validity}

For the considerations of this paper to be reliable, we must ensure that the bulk action is being used in its region of validity. This can be determined if we regard this bulk action as being obtained from string theory (appropriately dimensionally reduced). In that case, 
a first condition arises from string coupling: this condition is essentially the same as the large $N_c$ condition that applies in standard AdS/CFT correspondence. 
This is because the backgrounds we have considered do not involve the dilaton which is related to the string coupling. 
A second condition arises from higher derivative corrections: 
since the bulk electric field and hence the bulk curvature grows in the IR, we can restrict the region of validity by demanding that both curvatures, the field strength $F_{\m\n}2\pi\a'$ and the Ricci scalar $R\a'$ be small compared to the respective leading term in a string theoretic action. For the charged AdS/configurations, the former yields a restriction: 

\be
F_{\m\n}2\pi\a'=2\frac{\mu}{z_0 ^2} z (2\pi \a')<2\frac{\mu}{z_0} (2\pi\a')< P[g_{\mu\nu}]=\frac{L^2}{z^2} \implies \frac{4\p\mu z_0}{\sqrt{\l}} <<1
\ee
For a given value of $\l$, we obtain a restriction on the range of $\mu<\frac{290}{4\p} \sqrt{\l}$ MeV. Similarly, the gravitational part also gives an independent restriction
\be
2\pi\a' R=2\pi\a'\left(\z\frac{|F|^2}{4}+\frac{10}{3}\L\right)
<< 1
\ee
This condition depends on $\z$ and can be evaluated to

\be
\mu<\frac{1}{z_0}\sqrt{\left(20+\frac{\sqrt{\l}}{4\p}\right) \frac{9N_c}{\q N_f}}.
\ee
It seems clear that this is a weaker restriction than that coming from the gauge field.

When curvatures become large, string theory can be used to identify higher derivative correction terms with specific coefficients - for instance a DBI action for the electric field rather than the simple Maxwell action.

It is somewhat more difficult to quantify the errors being made by working with an IR incomplete geometry. Especially in view of the fact that we are including the effects of the IR geometry, at least partly, through the boundary conditions.

\section{Physical IR conditions}\label{PIR}
In an IR complete description of the finite density phase, we can expect a distribution of compact branes (and other solitons) representing baryons in the deep IR. This is motivated by studies \cite{Brandhuber:1998xy,Imamura:1998hf,Sakai:2004cn,Hong:2007kx,Kovensky:2021kzl} which show that such branes are dragged to the extreme IR of the (possibly warped) AdS geometries. Let us consider the (schematic) AdS path integral which gives the grand potential (pressure) broken up into the regions $\e<z<z_0$ and $z>z_0$
\be
Z=e^{-\b \O} = \int \mathcal{D} \psi e^{-S_{grav}[\psi]}=\int d\psi(z_0) \int^{\psi(z_0)} \mathcal{D} \psi e^{-\int^{z_0} _\e \mathcal{L}_{grav}[\psi]}\int_{\psi(z_0)} \mathcal{D} \psi e^{-\int^{IR} _{z_0} \mathcal{L}_{grav}[\psi]}  
\ee
where $\psi$ represents the collection of supergravity fields and $\psi(z_0)$ labels the boundary conditions on these fields at the IR cutoff $z_0.$ If the baryonic branes are localized behind the cutoff, the last term will capture their contribution. In our present work, we will model this contribution using an effective description. However, it must be emphasized that one can actually compute this part of the path integral perhaps along the lines of \cite{Kaplunovsky:2015zsa, Chakrabortty:2011sp, Jarvinen:2023jbr, Kovensky:2021kzl}. If the IR cutoff is sufficiently far away from the location of the baryon sources, we can expect that this IR physics feeds into the middle integral by supplying boundary conditions as we elaborate below. The middle term can then be calculated by using the saddle point approximation with the action we have discussed. Finally, the first integral over the boundary values will force us to minimize over those boundary values $\psi(z_0)$ that are unconstrained by the UV-conditions fixing the non-normalizable modes. 

To solve the equations for the fields in the region $\e<z<z_0$, we we will determine $g(z_0)$ and $\f_0$ in terms of the pressure $p_B$ and density $\rho_B$ computed from effective methods. There are two approaches to address this. We can choose $\phi_0$ so that the normalizable mode of the gauge field $Q(\mu,T)=\frac{12\p^2}{\q N_c N_f}\rho_B(\mu,T)$, and then determine the IR boundary condition $g(z_0)$ to ensure that the total pressure becomes
\be\label{pr}
p=p_B+\frac{N_c ^2}{8\pi^2}\frac{g_0+z_0 ^2}{z_0 ^6}
\ee
with $g_0$ the constant determined earlier.
Alternatively, we can set $g(z_0)=g_0$ and $\z Q^2z_0^2=p_B.$ This gives $Q=\sqrt{\frac{p_B}{\z z_0 ^2}}$ for the gauge field normalizable mode. Although this does not uniquely determine the boundary conditions, we will choose the former because if we envisage coupling the bulk gauge field to a scalar field, which could represent a condensate, setting $Q\simeq\rho_B$ leads to a natural coupling between the baryon {\em density} and the scalar field via the minimal coupling. 

In this work, we will use $p_B$, we will consider the following two effective descriptions: the first is the pressure of a gas of interacting baryons obtained in \cite{Fiorilla}. The second is the 2-flavor NJL model at finite temperature and chemical potential \cite{Klevansky:1992qe}. The reason for choosing the first equation of state is that at low density, the baryons maybe described as a gas of solitons on the world-volume of flavor branes. At somewhat larger densities, nuclear physics suggests the appearance of a quarkyonic phase which we model as a gas of constituent quarks whose pressure is modelled by the NJL model \cite{Klevansky:1992qe}. However, since \cite{Fiorilla} do not include the strange quarks in their analysis, we use the two flavor NJL model as the higher density phase rather than the three flavor version. 

Note that $p_B$ is suppressed by $\frac{N_f}{N_c}$ compared to the second term in \eqref{pr} which is consistent with the idea that it is generated by flavor degrees of freedom (arising from embedded branes, for instance). Even so, the holographic identity relating the energy density computed from the normalizable modes of the metric to  the derivative of the pressure will not be satisfied. 

\subsection{van der Waals gas}\label{vdw}
The pressure for symmetric nuclear matter system  was computed using chiral effective field theory in \cite{Fiorilla}.  In this work, the authors used two and three body contact interactions and also included the effects of the $\D$ resonance in intermediate states. They found that the pressure has a van der Waals form at temperatures below 20 MeV. 

At zero temperature, we can take
\be
p_{vdW}=-\frac{K\r^2}{9\r_0}\left(1-\frac{\r}{\r_0}\right)
\ee 
where $\rho$ is the baryon number density in units of fm$^{-3}$ and using 
$p(\r,T)=\r^2\frac{\del \Bar{F}(\r,T)}{\del\r}$
we obtain the Free Energy per particle $\Bar{F}(\r)$,
\be \label{fe}
\Bar{F}=\frac{K}{2}\left(\frac{\r^2-2\r \r_0}{9\r_0^2}\right) + c
\ee
The energy per particle $\Bar{E}=\Bar{F}(\r,0)$ near nuclear saturation density ($\r_0$) up to the second order 
\be 
\Bar{E}\sim m_B+\bar{E_0} + \frac{K}{2}\left(\frac{\r-\r_0}{3\r_0}\right)^2
\ee 
and matching with \eqref{fe}, we get $c=m_B$ and $\bar{E_0}=-\frac{K}{18} \approx -16$ MeV is the binding energy of nuclear matter at saturation density. The chemical potential at T=0:
\be 
\m=m_B+\bar{E_0}+ \frac{K}{2}\left(\frac{\r-\r_0}{3\r_0}\right)^2-\frac{\r K}{9\r_0}\left(1-\frac{\r}{\r_0}\right)
\ee  
When $\r<\rho_0$, by the Maxwell construction, the baryon contribution to the pressure is taken to be identically zero and the chemical potential becomes constant at $\mu=923$MeV.

In this van der Waals form, in contrast to \eqref{pcads}, the pressure vanishes at zero density because the pion gas and glueball contributions have not been included in the computation. 

\begin{figure}[h]
    \centering
    \subfigure[Pressure vs $\r$]{\includegraphics[width=0.3\textwidth]{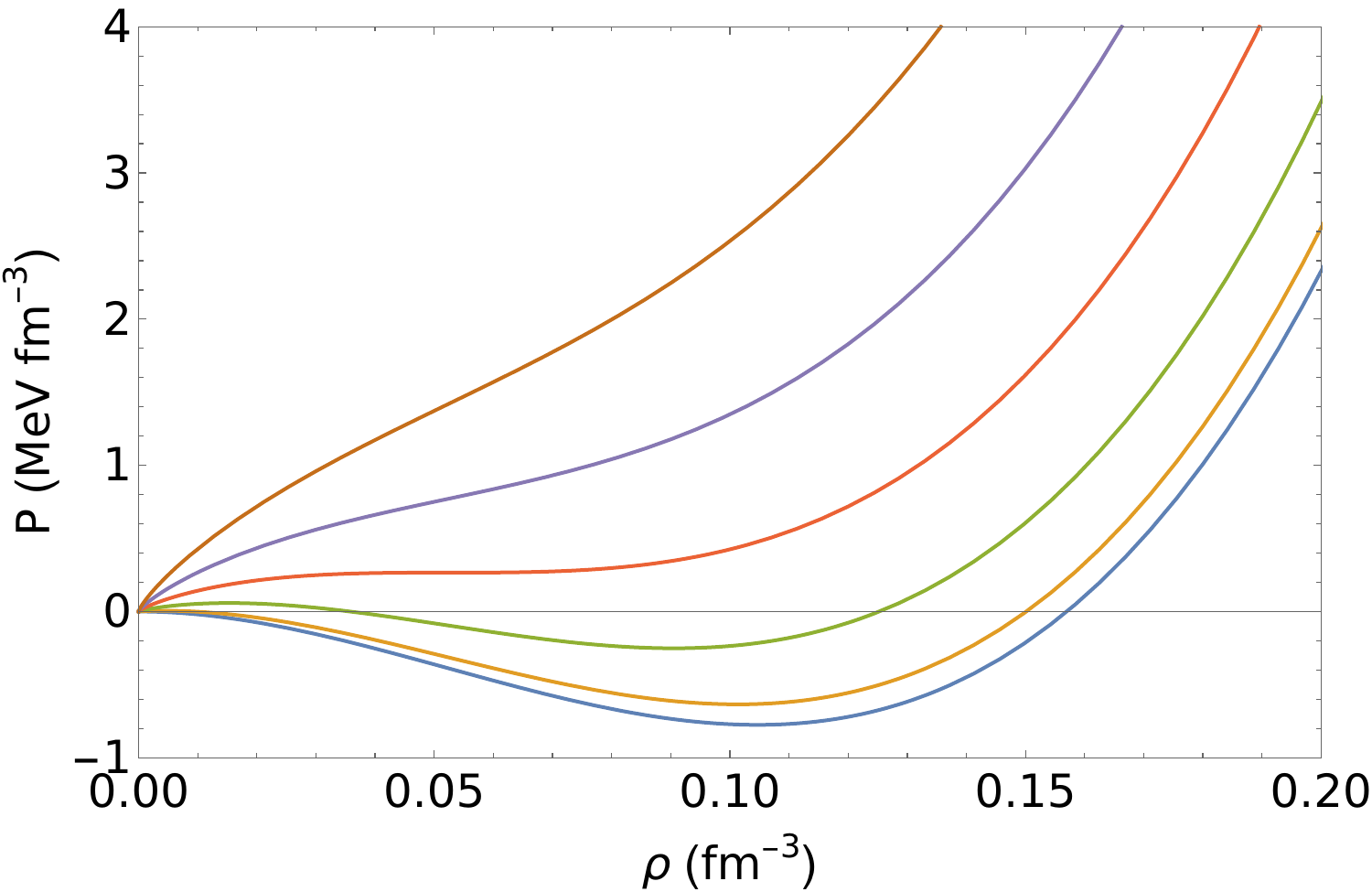}}
    \subfigure[Free energy pp vs $\r$]{\includegraphics[width=0.3\textwidth]{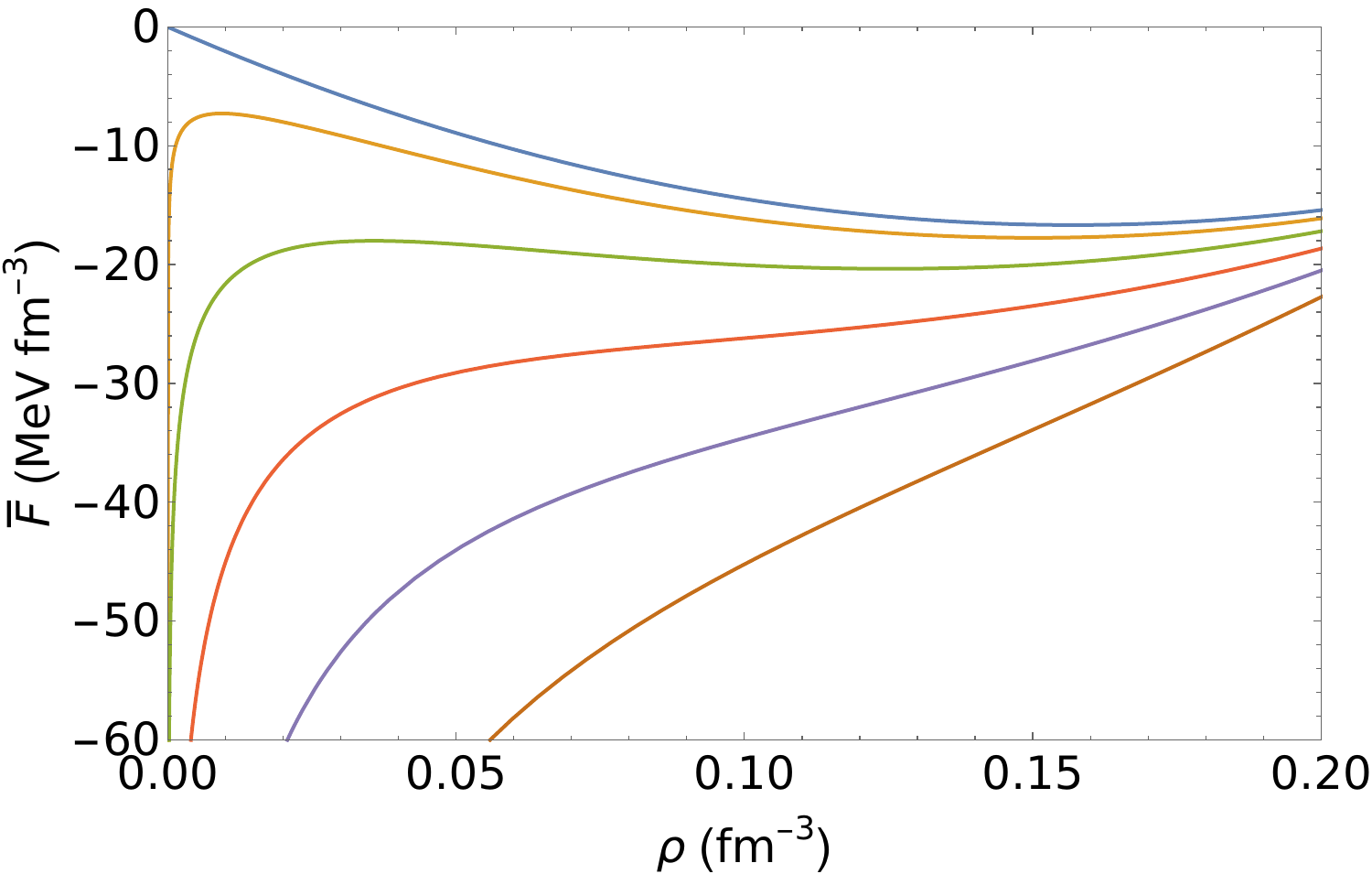}}
    \subfigure[Pressure vs $\m$]{\includegraphics[width=0.3\textwidth]{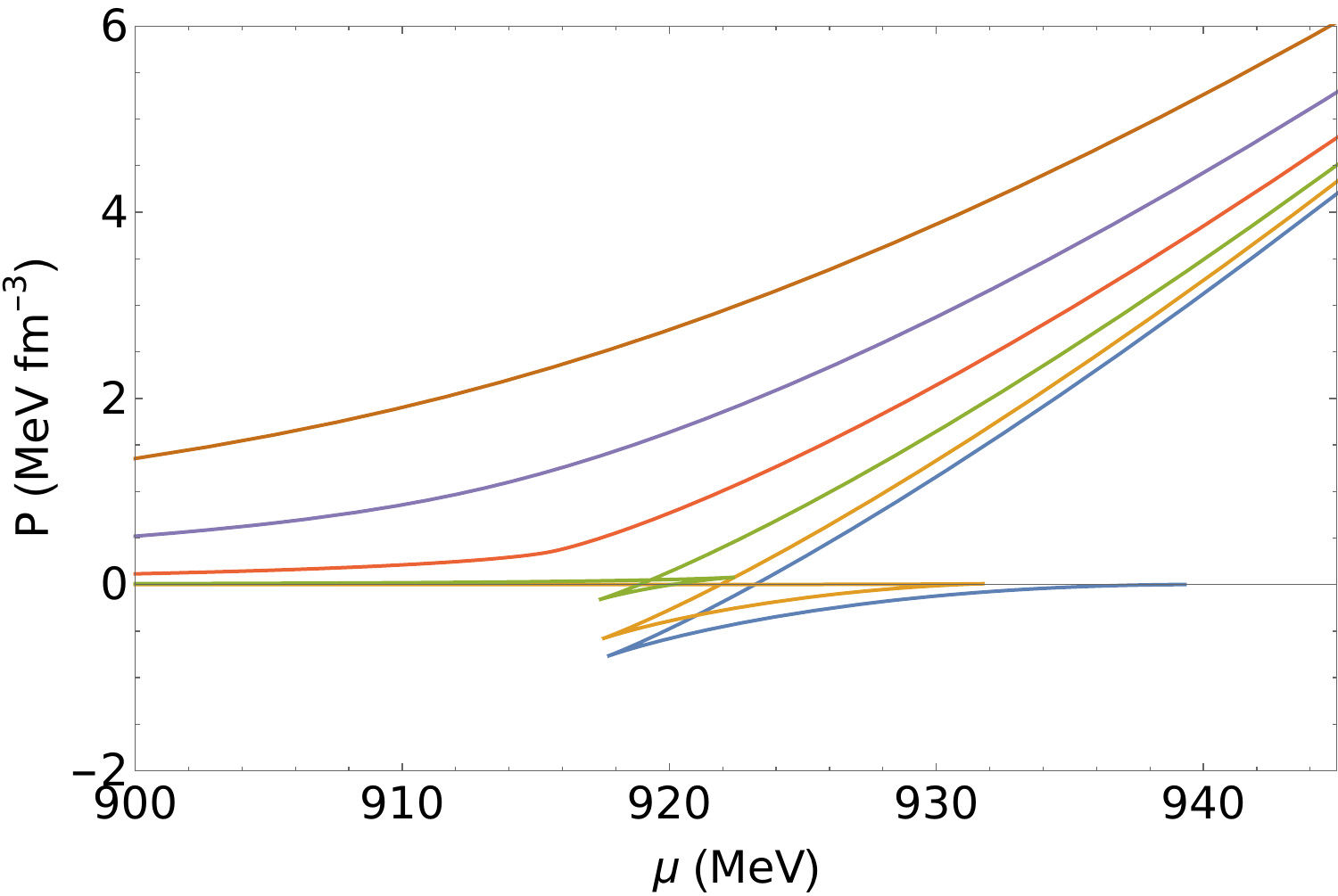}}
    \caption{vdW at finite temperature}
    \label{fig:vdw}
\end{figure}

At finite temperature, the following seems a reasonable fit to the graphs of \cite{Fiorilla}
\bea
p &=&p(0)+a T^2 \left(\frac{\rho }{1-b \rho }\right)^{4/5}\\
\mu &=& \m (0)-0.1 T+ T^2 \left(0.167\, -5 a \left(\frac{\rho }{1-b \rho }\right)^{-1/5} +\frac{a}{\rho }\left(\frac{\rho }{1-b \rho }\right)^{4/5}\right)
\eea
with $b=1.92$, $a=0.027.$ In these formulae, $\rho_0=0.157$ fm$^{-3}$ and  $K=300$ MeV. While the applicable range of density is $0<\rho<\frac{1}{b} \sim 3.3\rho_0$ which can be translated into the range $0<\frac{\mu_B-m_B}{N_c}<325$  MeV (at $T\simeq 1$MeV). This fit which is depicted in figure \ref{fig:vdw} is  adequate since we are only interested in demonstrating a proof of principle in hardwall AdS holography.

\subsection{Quarkyonic Liquid system}
As the density increases, it is widely regarded that the baryonic liquid of the preceding section melts into a liquid of strongly interacting quarks which is termed the `quarkyonic liquid'. Since strange quark effects were not considered in subsection \ref{vdw}, it is appropriate to use the 2-flavor NJL. The 2-flavor NJL model gives a pressure and energy density of the form \cite{Klevansky:1992qe},\cite{Asakawa:1989bq}
\bea\nn\label{njlP}
p_B=\frac{N_f N_c}{\p^2} \int^\L dp\: p^2\left( E_p + T\log\left[(1+e^{-\b(E_p+\m')})(1+e^{-\b(E_p-\m')})\right]\right)\\+ \frac{G}{2 N_c}n^2-G \left(1+\frac{1}{4N_c}\right) \sigma ^2
\eea
\bea\nn\label{njle}
\e_B=\frac{N_f N_c}{\p^2} \int^\L dp \:
 p^2 E_p \left(\frac{1}{1+e^{\b(E_p-\m)}}+\frac{1}{1+e^{\b(E_p+\m)}}-1\right)\\+ \frac{G}{2 N_c}n^2+G \left(1+\frac{1}{4N_c}\right) \sigma ^2
\eea
where $E_p=\sqrt{p^2+M^2}$ is the energy of {\em constituent quarks} 
and, the chiral condensate and baryon number density are defined by:
\bea
\s&=&-\frac{N_f N_c}{\p^2} M \int^\L dp \frac{p^2}{E_p}\left(1-\frac{1}{1+e^{\b(E_p-\m')}}-\frac{1}{1+e^{\b(E_p+\m')}}\right)\\
n&=&\frac{N_f N_c}{\p^2}\int^\L dp\  p^2 \left(\frac{1}{1+e^{\b(E_p-\m')}}-\frac{1}{1+e^{\b(E_p+\m')}}\right)
\eea
At zero temperature and chemical potential, the energy density is negative $\e_B\sim -\L^4 + \L^2$. Here, we have used the fact that $G\L^2\sim\mathcal{O}(1)$. Therefore, we subtract the zero point pressure and energy density from \eqref{njlP} and \eqref{njle}, respectively.

The constituent quark mass $M$ depends on the temperature and chemical potential and the UV cutoff $\L$ and is determined by solving the self-consistency conditions
(or gap equations):
\be
M=m_0-2 G \left(1+\frac{1}{4N_c}\right)\s 
\qquad \mu'=\mu-\frac{G}{N_c}n
\ee
where $m_0$ is the current quark mass.
We numerically solve these self-consistency conditions and determine all thermodynamic quantities. We use the same parameter choices mentioned in \cite{Klevansky:1992qe}
\[\L=631 \text{ MeV} \qquad G\L^2=2.02 \qquad m_0=5.5 \text{ MeV}\]

Thus, we are regarding this higher density phase as quarkyonic system with density dependent masses.
\begin{figure}[h]
    \centering
    \includegraphics[width=0.45\textwidth]{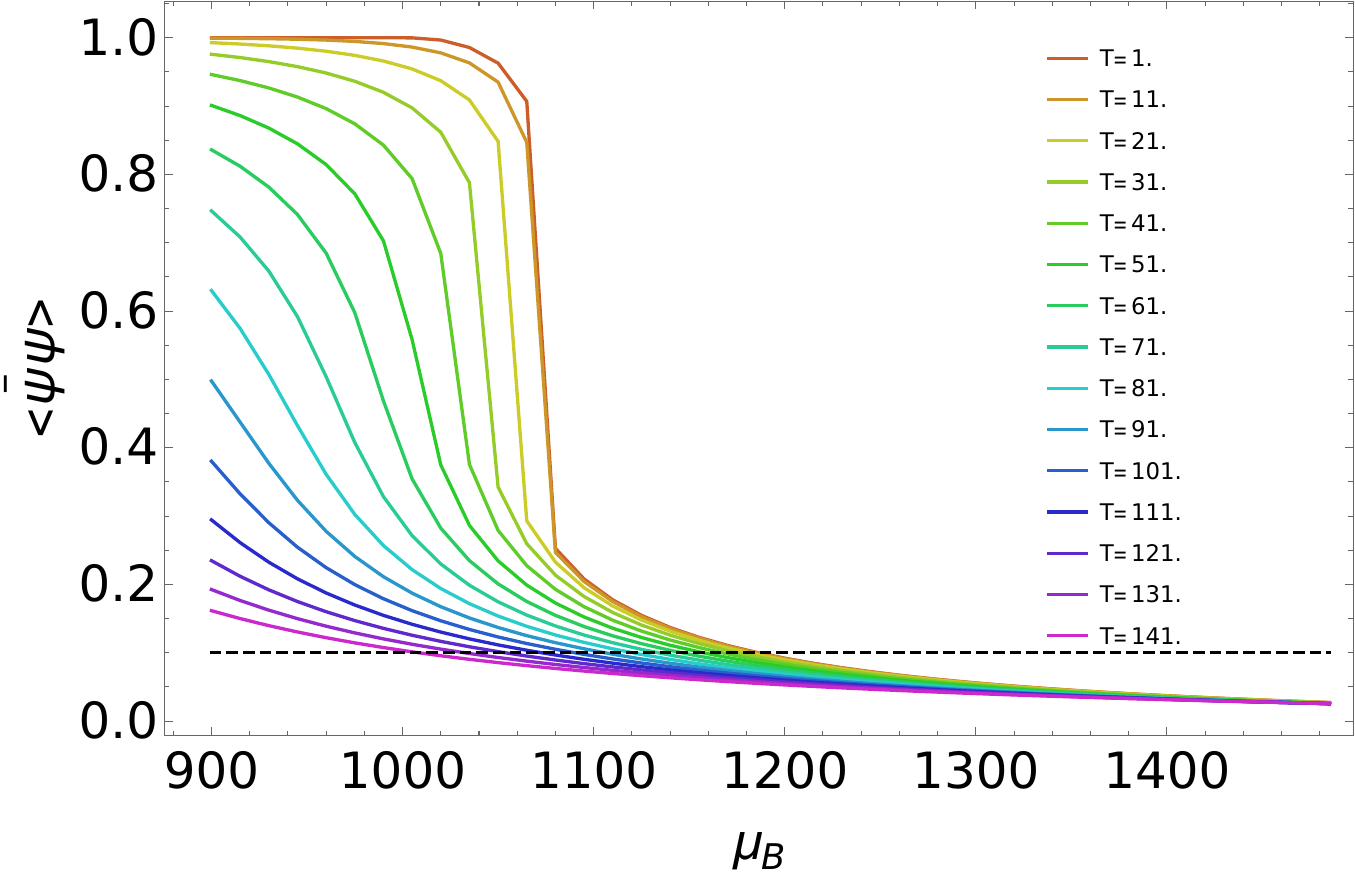}
    \caption{Reduced chiral condensates for various temperatures.}
    \label{fig:ccond}
\end{figure}
At $T=0$, chiral symmetry is nearly restored at around $\mu_B=1200$ MeV as depicted in figure \ref{fig:ccond} where we have normalized it by its value at zero temperature and density. For larger temperatures, the chiral condensate goes to zero much slower but above $T=140$ MeV, the chiral condensate remains quite small. 
\begin{figure}[h]
    \centering
    \subfigure[Pressure]{
    \includegraphics[width=0.45\textwidth]{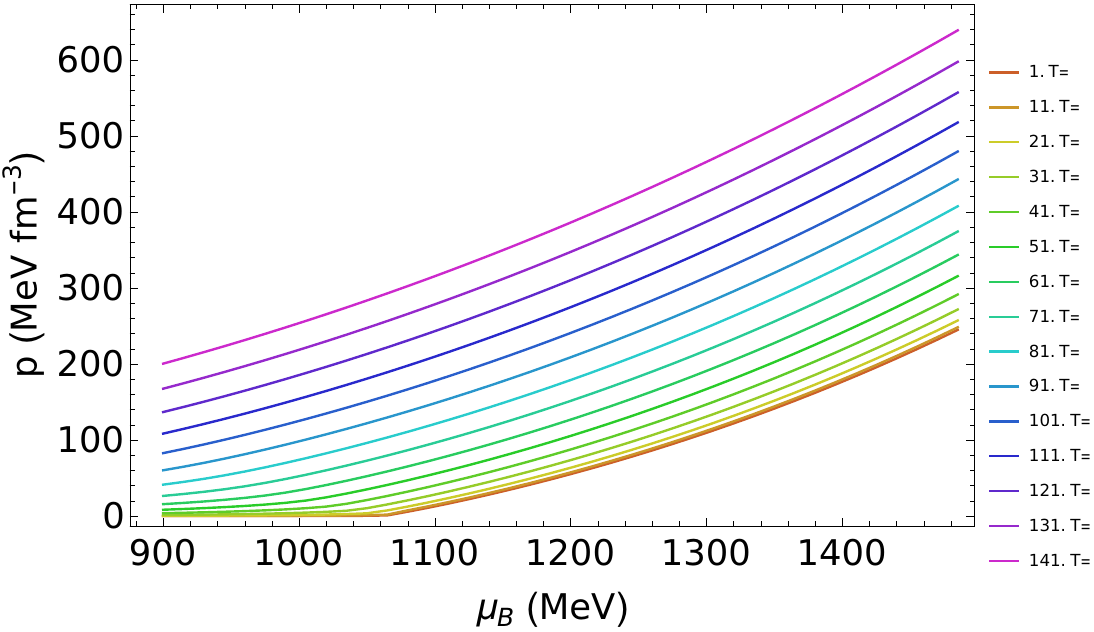}}\hfill
    \subfigure[Energy density]{
    \includegraphics[width=0.45\textwidth]{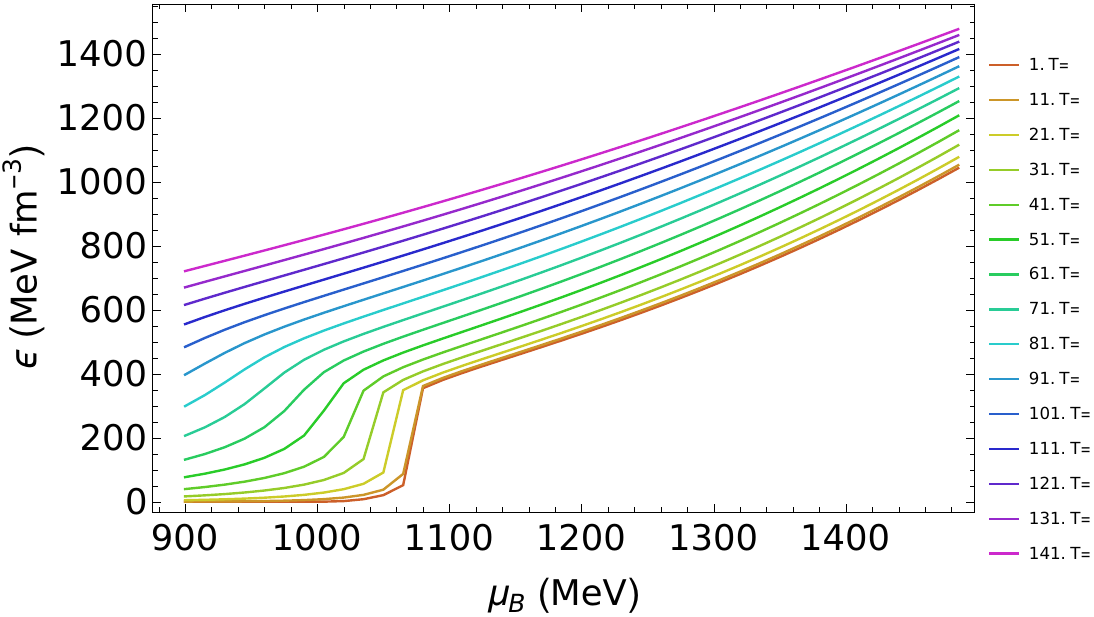}}
    \caption{Pressure and energy density as functions of chemical potential for different values of temperature in the NJL model}
    \label{fig:pnjl}
\end{figure}
The graph of the pressure and energy density is shown in figure \ref{fig:pnjl}. The energy density graph also shows a jump. 
\begin{figure}[h]
   \centering
    \subfigure[Equation of state]{
    \includegraphics[width=0.32\textwidth]{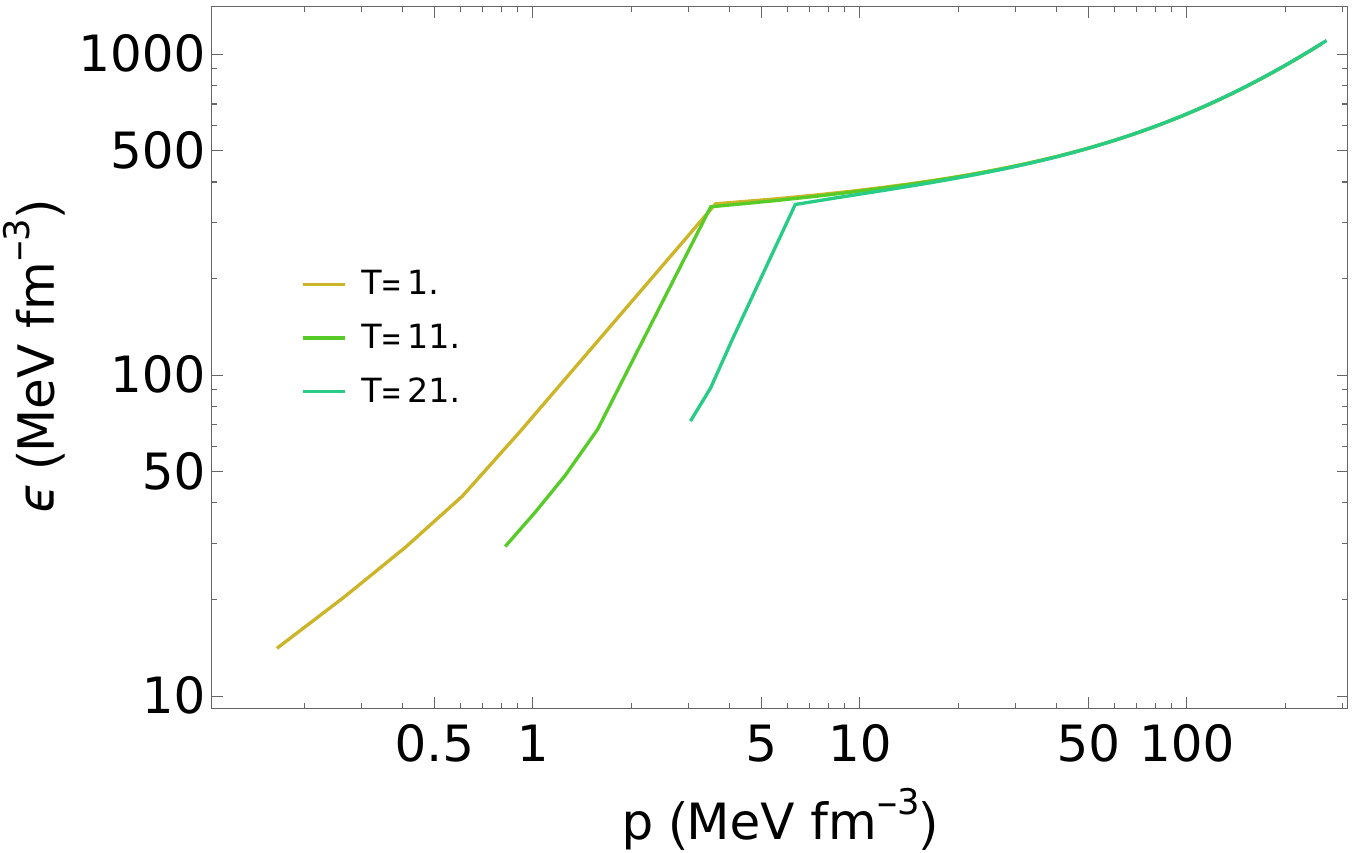}\label{fig:eVSp}}\hfill
    \subfigure[Number density]{
    \includegraphics[width=0.32\textwidth]{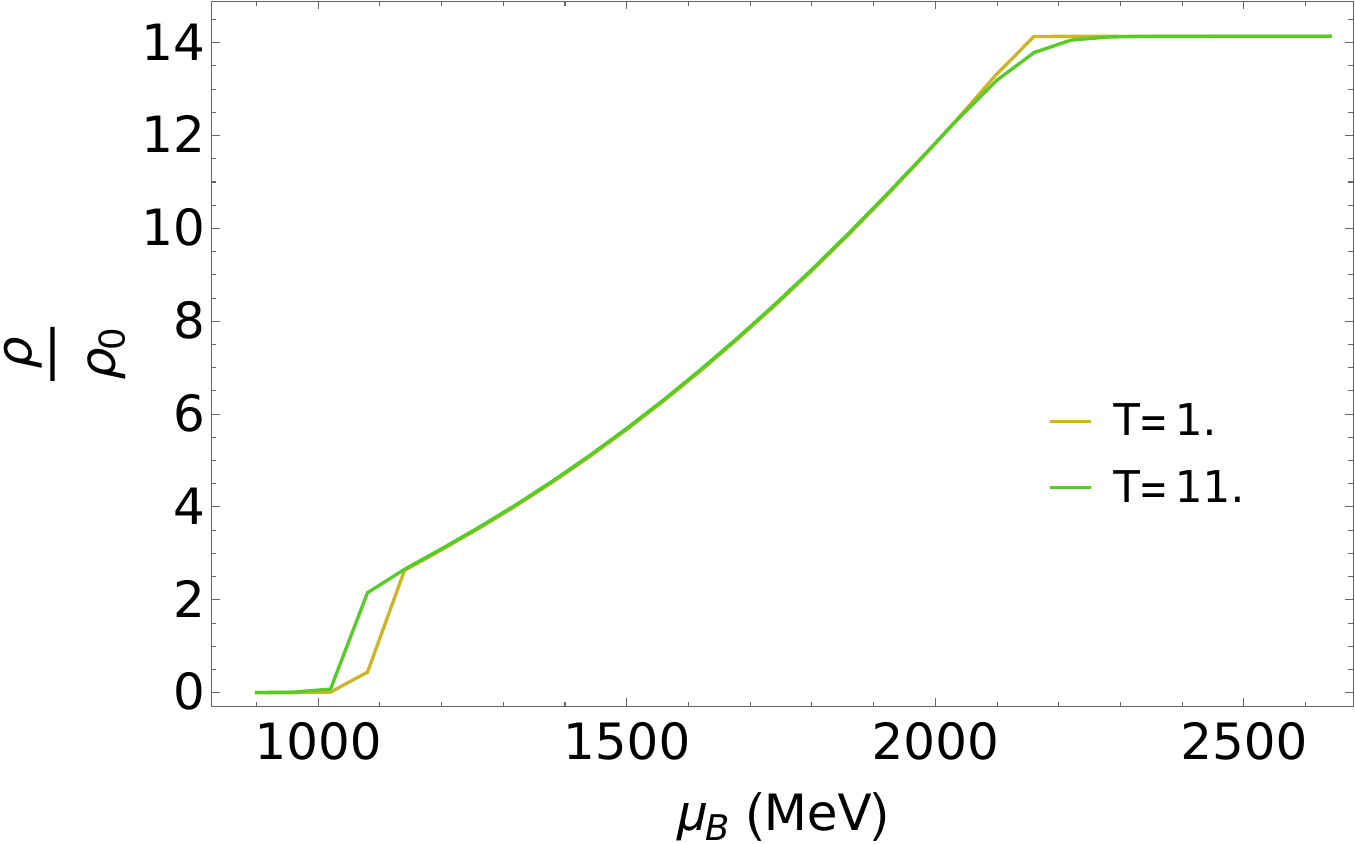}\label{fig:NdVSmu}}\hfill
    \subfigure[Speed of sound]{\includegraphics[width=0.3\textwidth]{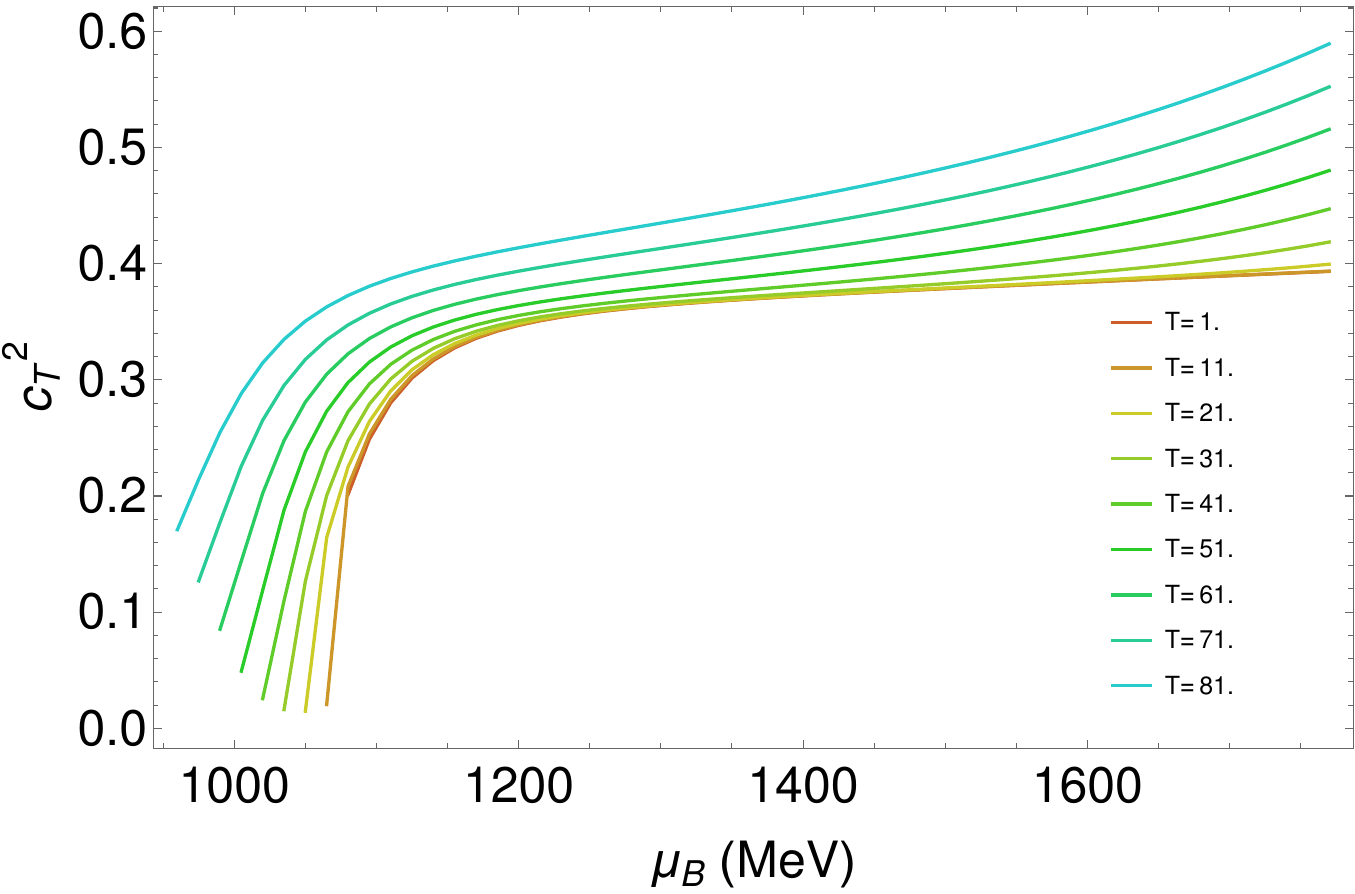}\label{fig:cVSmu}}
    \caption{Energy density as a function pressure, number density and speed of sound squared as functions of chemical potential at fixed temperature for NJL model}
    
\end{figure}
Plotting the energy density as function of pressure clearly shows a change in behaviour. At low temperatures, for small chemical potential (or pressure), $\e\sim p$. However, at large chemical potential, $\e\sim \sqrt{p}$. The exact equation of state depends on the temperature and can be extracted using a fitting function.  For example, when $T=1$, the $\e(p)$ graph maybe approximated using two pieces:   

\begin{equation*}
 \e \simeq   \begin{cases}
        59.7 p + 8.97 \sqrt{p}  & \text{if } p < 5 \\
             308 + 15.8 \sqrt{p} + 1.81 p  & \text{if } p \ge 5
            \end{cases}
\end{equation*}
The appearance of the $\sqrt{p}$ term is interesting to holographists in the sense that it naturally arises from the action of the D-brane \cite{Karch:2007br,Hoyos:2016zke}. 
The figure \ref{fig:NdVSmu} shows that the number density saturates at large chemical potentials. The figure \ref{fig:cVSmu} shows the speed of sound squared $c_T^2$ computed at the constant temperature as a function of chemical potential. 
\[c_T^2=\frac{\del p}{\del \e}\Big|_T\]
It can be seen that for small temperature, the speed of sound approaches the conformal value $c_T^2\approx\frac{1}{3}$. However, at high temperatures and chemical potential, it can even break the causal limit $c_T^2 > 1$. 

\subsection{Phase diagram}
We will first show that the quarkyonic phase will dominate at high densities over the baryonic van der Waals liquid. 

In comparisons of this kind, we must always compare the two systems at the same temperature and chemical potential. This is because gradients in temperature and chemical potential drive non-equilibrium transport. Thus, if the chemical potential varies in a medium, mass transport will occur leading to a homogenization of $\mu_B$.

\begin{figure}[t]
\centering
    \subfigure[T=1 MeV]{\includegraphics[width=00.3\textwidth]{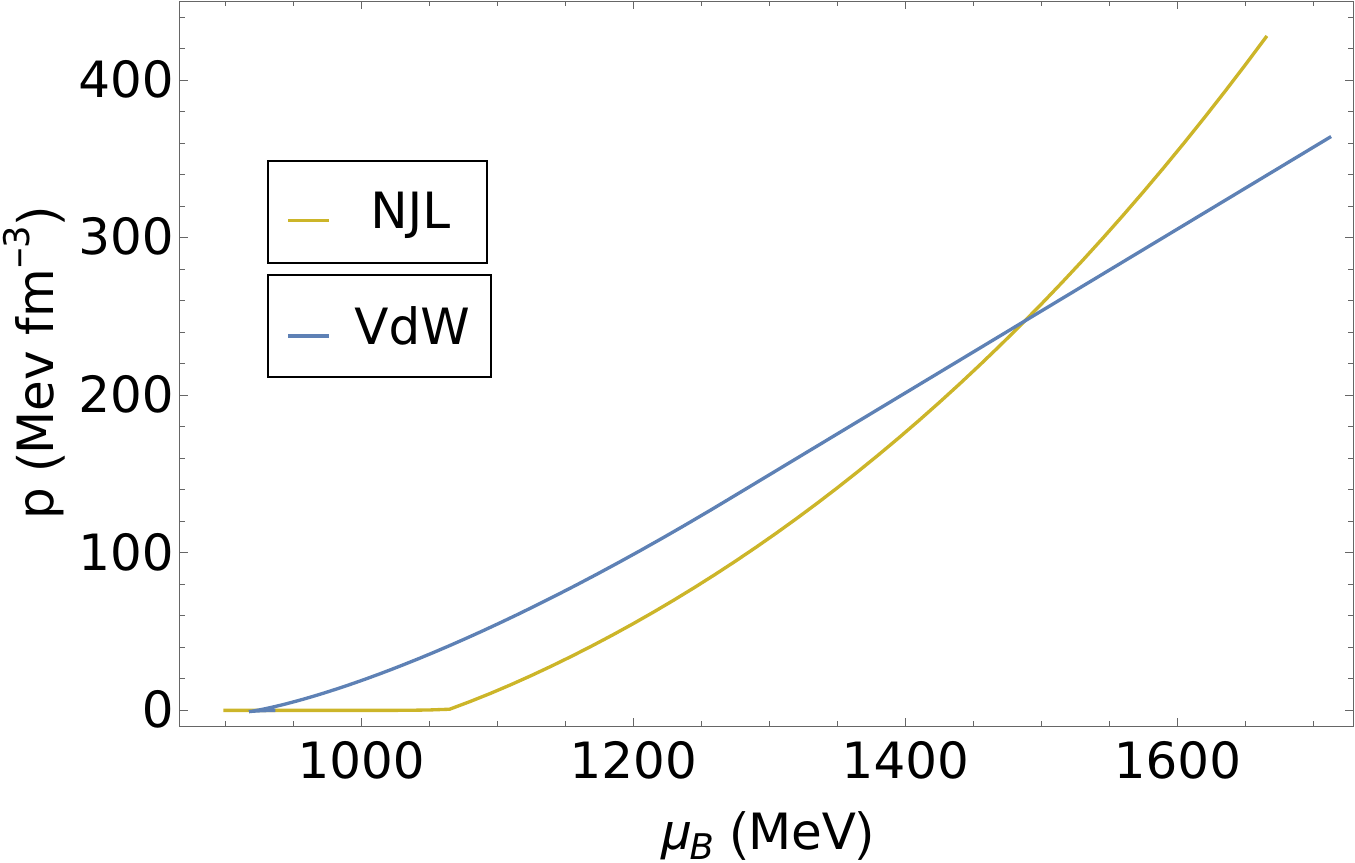}}
    \subfigure[T=15 MeV]{\includegraphics[width=0.3\textwidth]{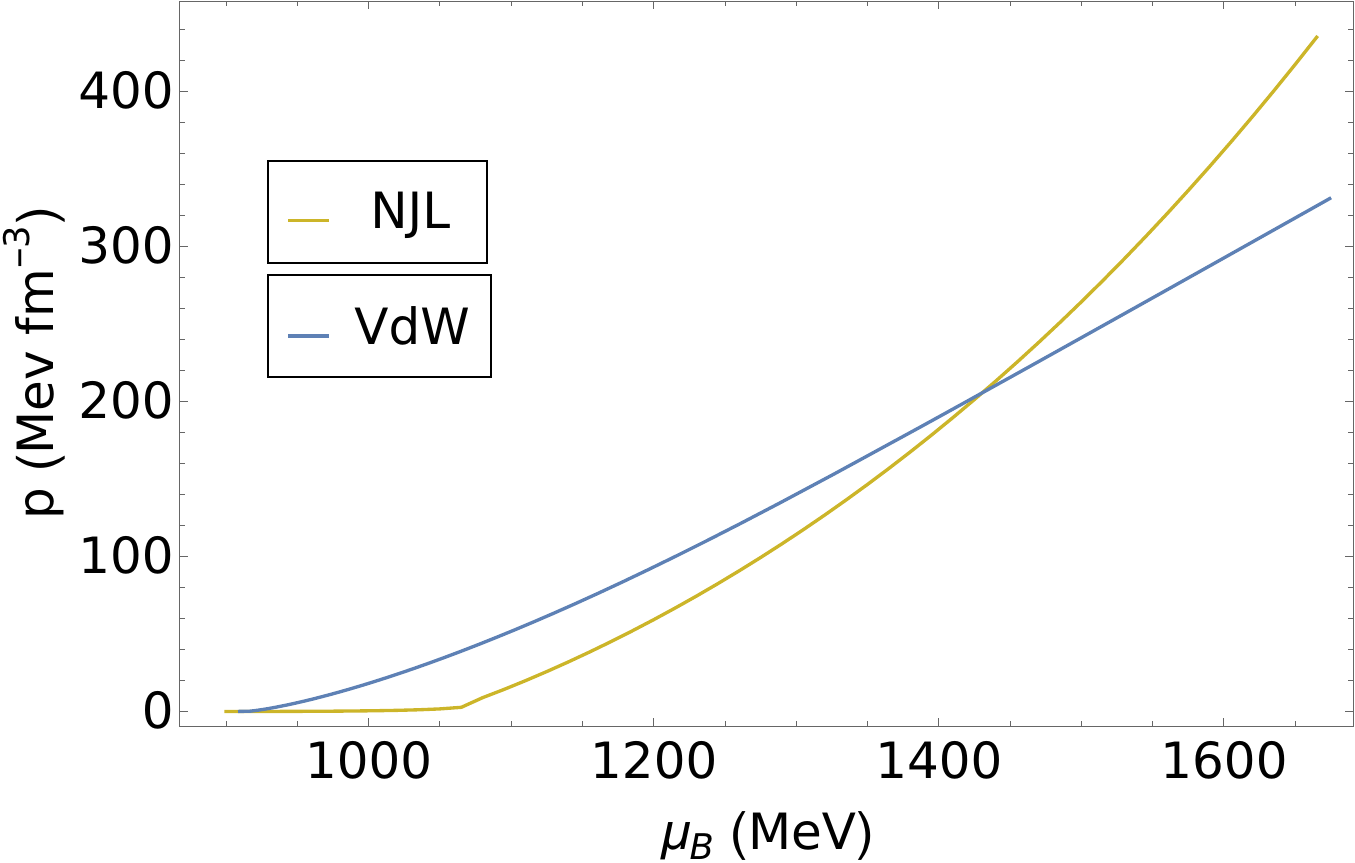}}
    \subfigure[T=30 MeV]{\includegraphics[width=0.3\textwidth]{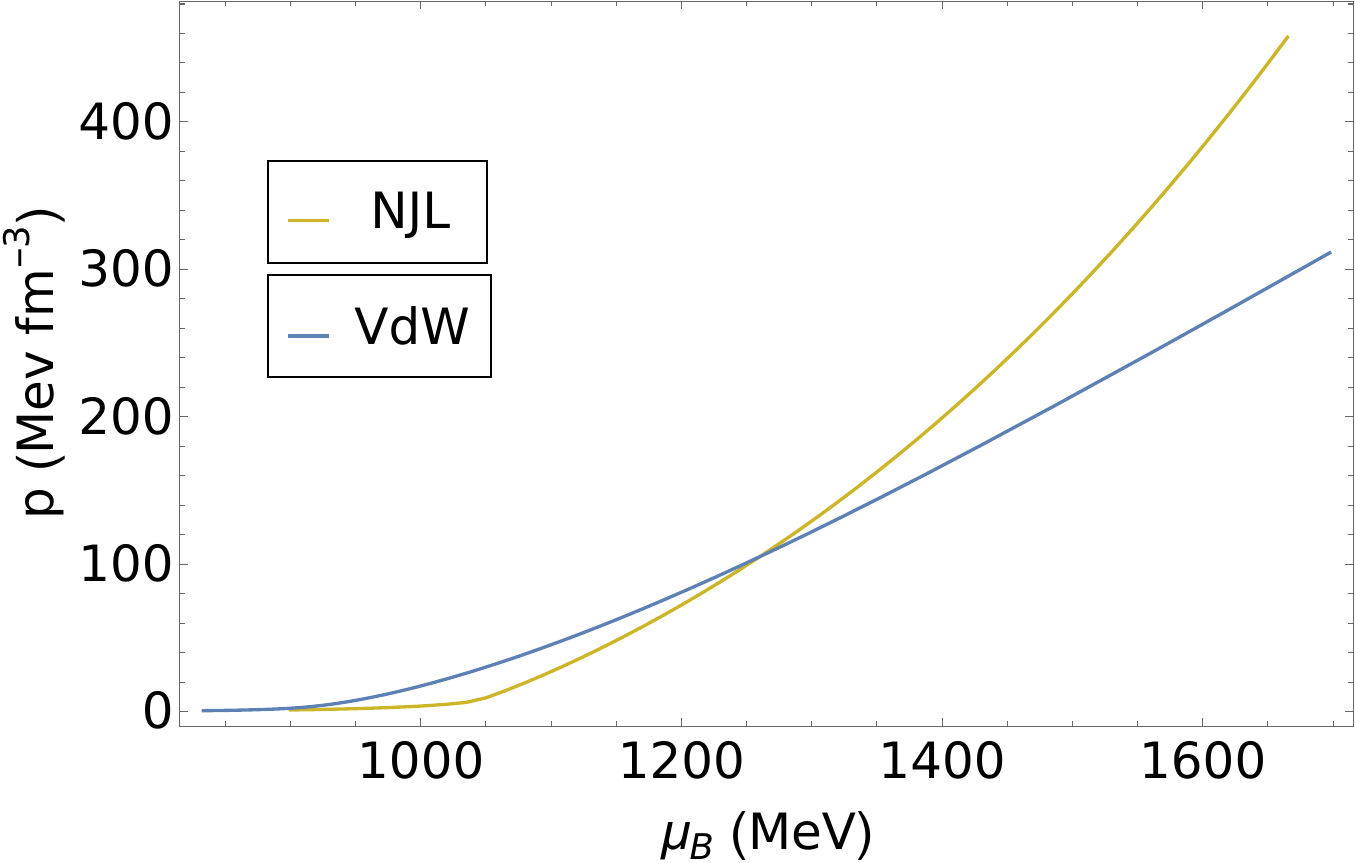}}
    \caption{vdW to NJL transition at low temperatures}
    \label{fig:vdWNJLT}
\end{figure}

We will first compare the pressures of the van der Waals gas-liquid with that computed from the 2-flavor NJL model at zero temperature. This is because these are descriptions of the same phase using different effective degrees of freedom. The better description will be that which leads to higher pressure and we may expect that the NJL model is more appropriate at higher densities.
In figure \ref{fig:vdWNJLT},  we plot the pressures of the van der Waals gas using the fit discussed in subsection \ref{vdw} alongside those obtained using the 2-flavor NJL model for a few corresponding values of the temperature.
 
We see that the NJL description of the baryonic system as made of a gas of constituent quarks - i.e., a `quarkyonic' description must be the correct description of the system at some higher density. At low density, we can get larger pressure by ``packing" the quarks into baryons and baryons into clusters of baryons. As the temperature increases, the quarkyonic phase dominates the nucleonic description at a lower value of $\m_B.$ 

We will next compare the pressure of the NJL gas with the extremal charged black hole representing the deconfined phase. In this case, we will need to fix the constant $g_0$ which represents the glueball contribution to the pressure and also the parameter $\z.$ 
\begin{figure}[t]
    \centering
    \subfigure[$\z=0.75$]{
\includegraphics[width=0.45\textwidth]{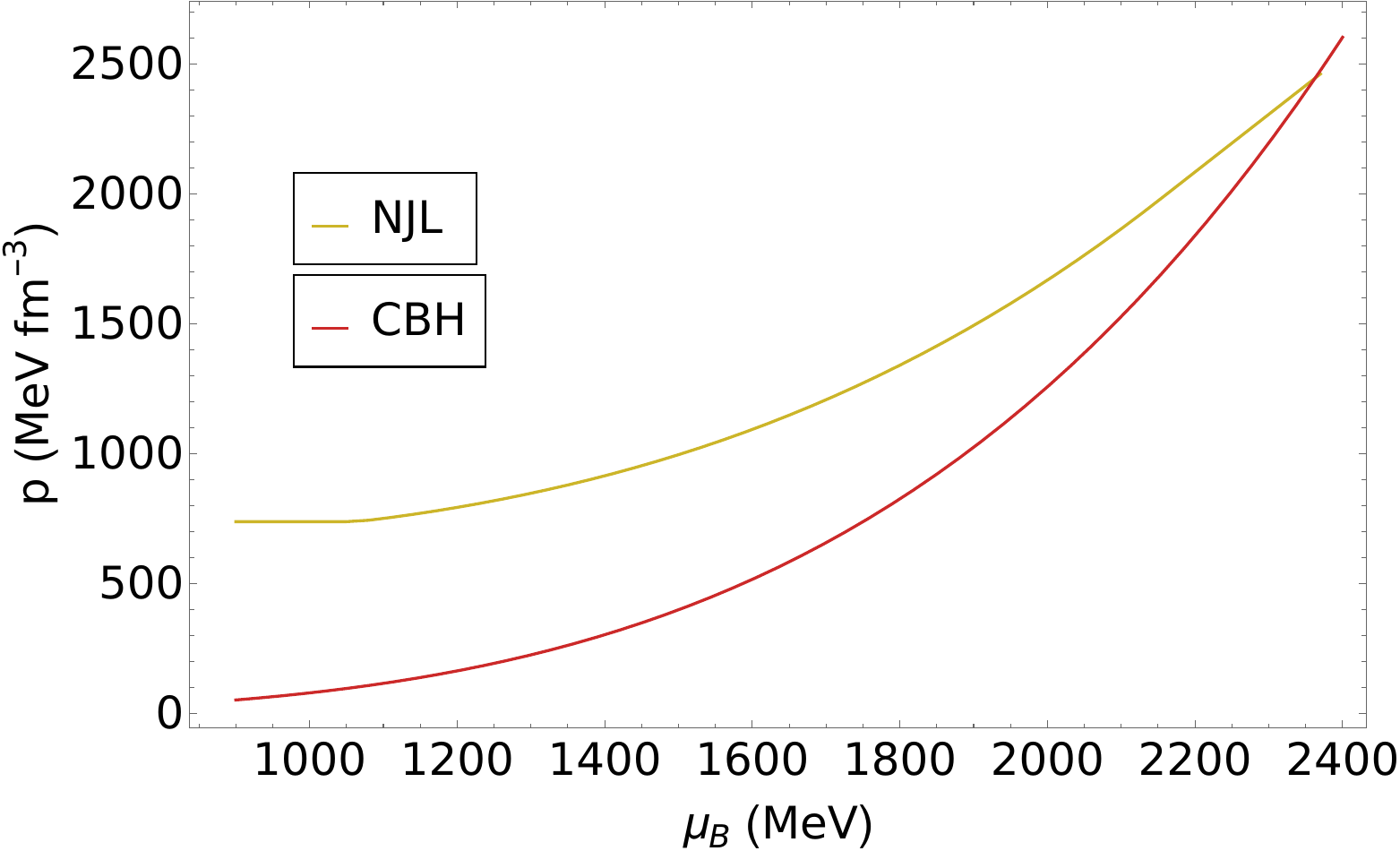}}
\subfigure[$\z=0.77$]{
\includegraphics[width=0.45\textwidth]{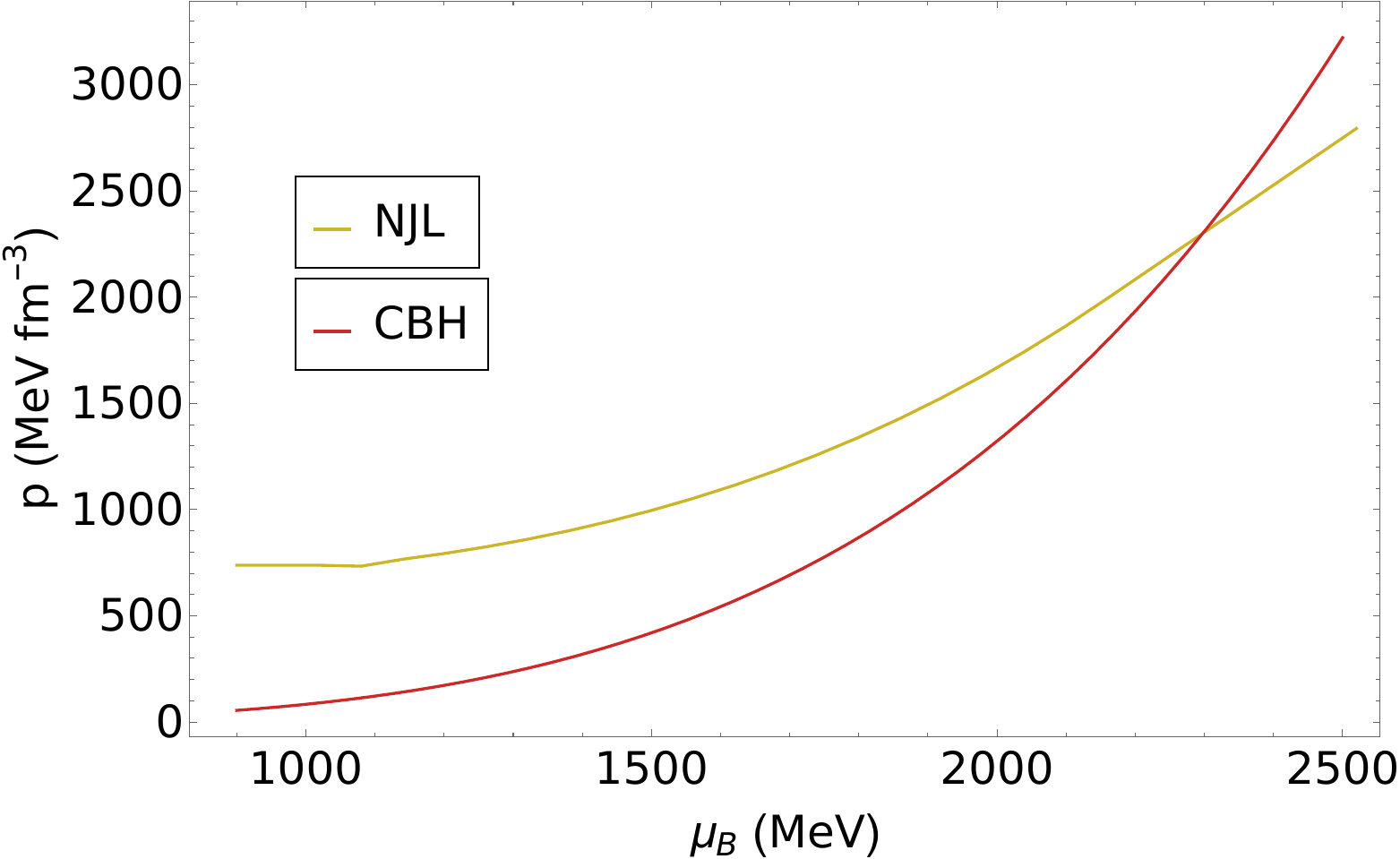}}
\subfigure[$\z=1$]{
\includegraphics[width=0.45\textwidth]{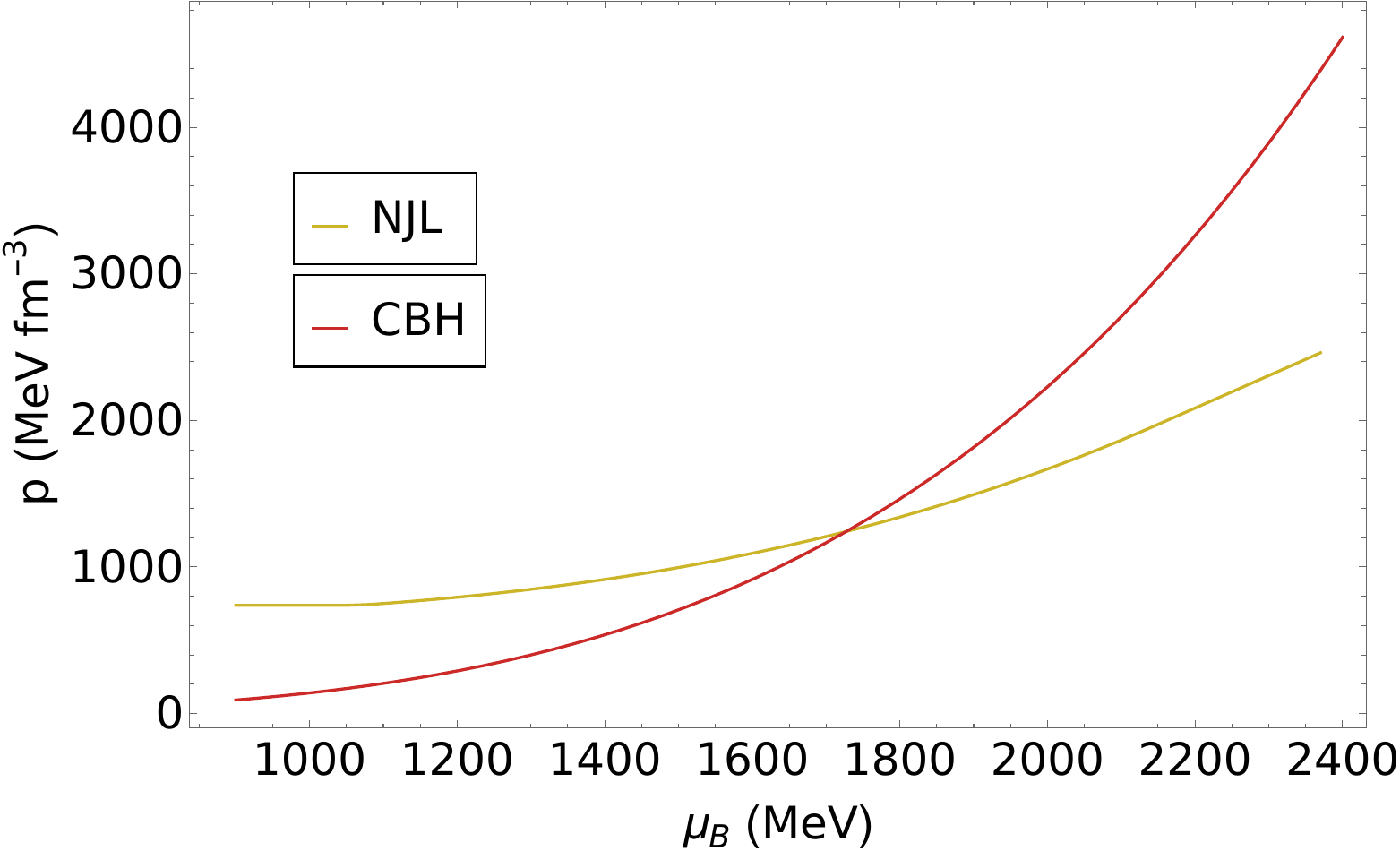}}
\subfigure[$\z=1.25$]{
\includegraphics[width=0.45\textwidth]{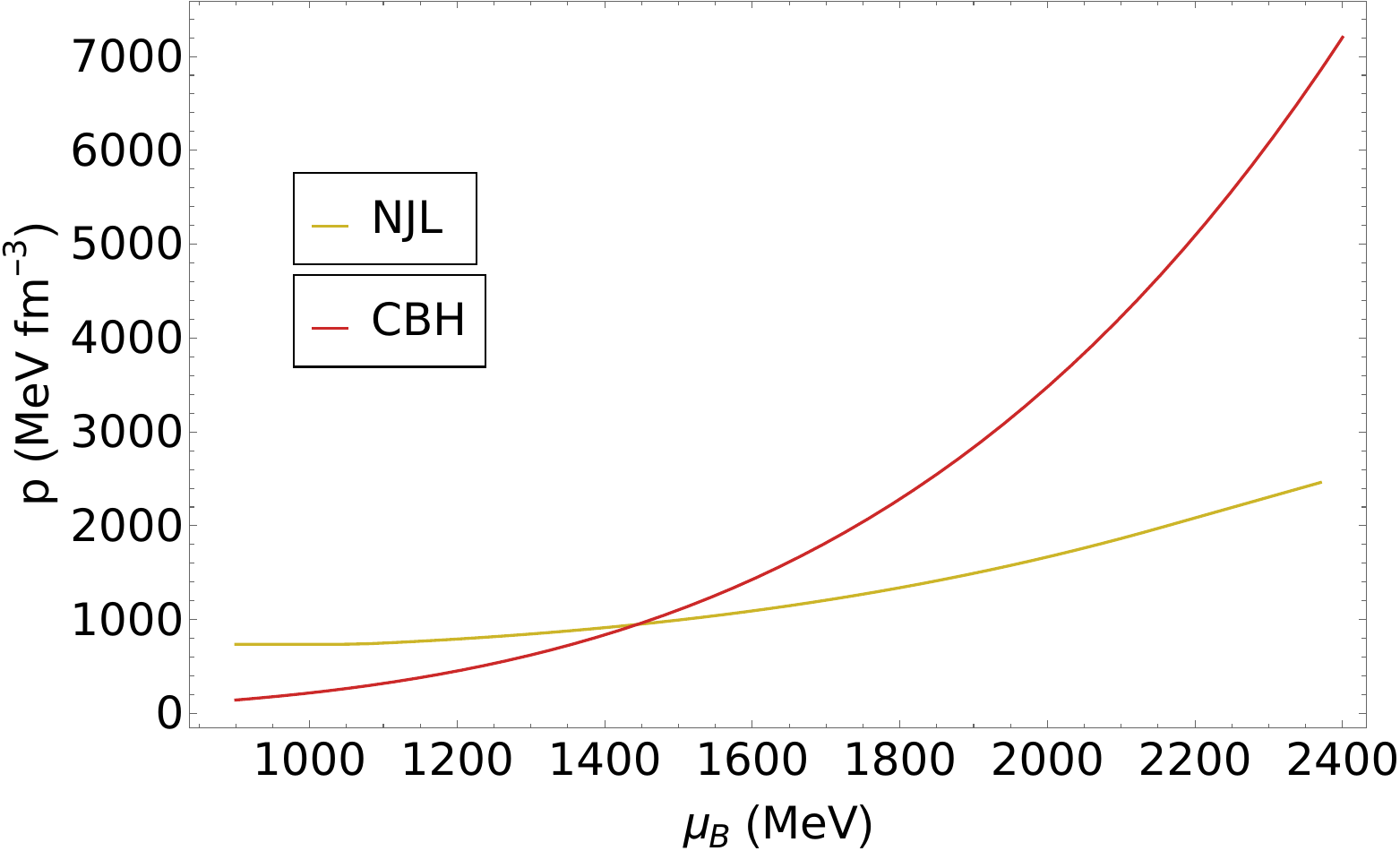}}

    \caption{Deconfinement transition at T=0}
    \label{fig:Aguirre-PT}
\end{figure}
We observe that for the given choices of parameters $z_0$ and $g_0$, for $\z\lesssim 0.75$, the deconfined (CBH) phase consistently exhibits lower pressure than the Quarkyonic (NJL) phase, making the NJL phase the preferred choice across the chemical potential ranges expected in this QCD phase diagram. Conversely, for $\z\gtrsim 1.25$, there exists a threshold beyond which the NJL phase is never the preferred choice across any chemical potential range. 
In figure \ref{fig:Aguirre-PT}, we see that the charged black hole pressure (for $\z=1, g_0=6$) exceeds that of the Quarkyonic NJL model at around $\m_B \simeq1800$ MeV. We emphasize that this is a {\em prediction} of AdS/Holography. Since the black hole phase is associated with a Polyakov loop having a nonzero expectation value \cite{Colangelo:2010pe}, this represents a deconfinement transition. At these values of the chemical potential, the chiral condensate is nearly zero. This agrees with the recent work of \cite{Costa:2019bua}. 
It must be pointed out that this is different from what happens at low chemical potential and finite temperature. In the PNJL model \cite{Costa:2019bua} and in  \cite{Singh:2022obu}, we see that the chiral condensate does not quite vanish even past the deconfinement transition (although the holographic system shows a rather small logarithmic condensate).

\begin{figure}
    \centering
    \subfigure[ $\zeta=0.77$, $g_0=6$ ]{\includegraphics[width=0.45\textwidth]{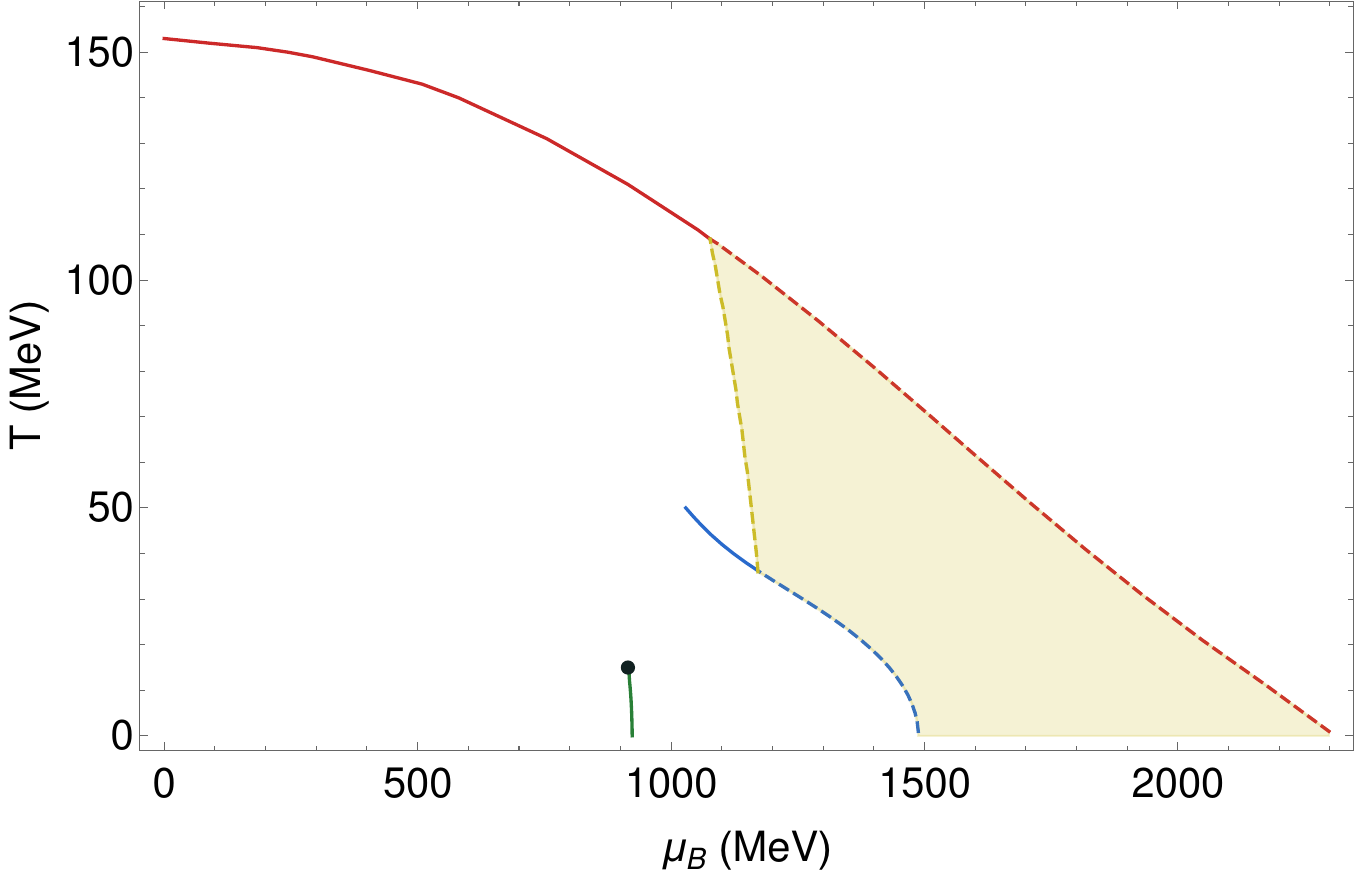}}\hfill
    \subfigure[$\zeta=1$,  $g_0=6$]{\includegraphics[width=0.45\textwidth]{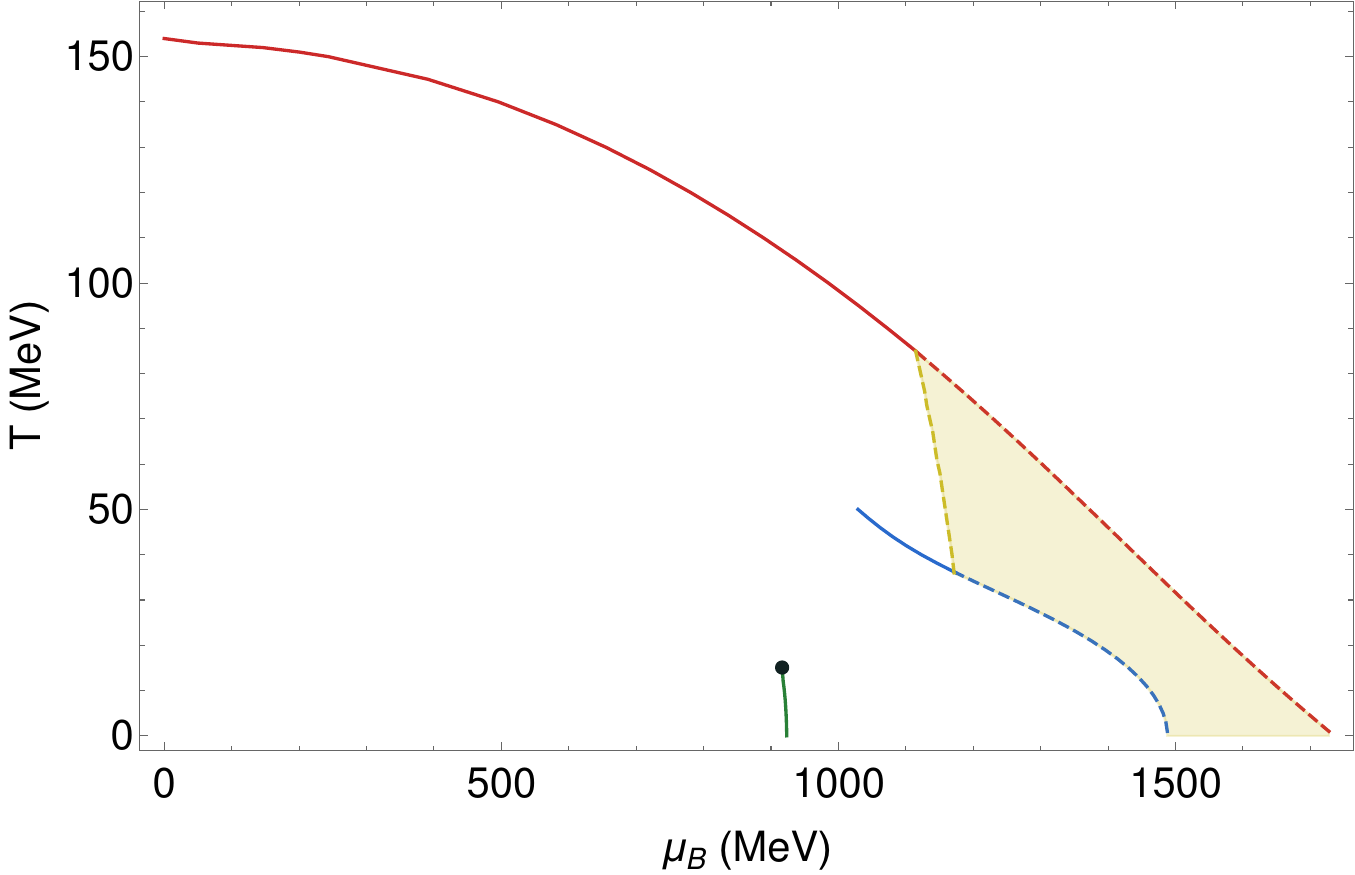}}
    \subfigure[$\zeta=0.77$, $g_0=27 $]{\includegraphics[width=0.45\textwidth]{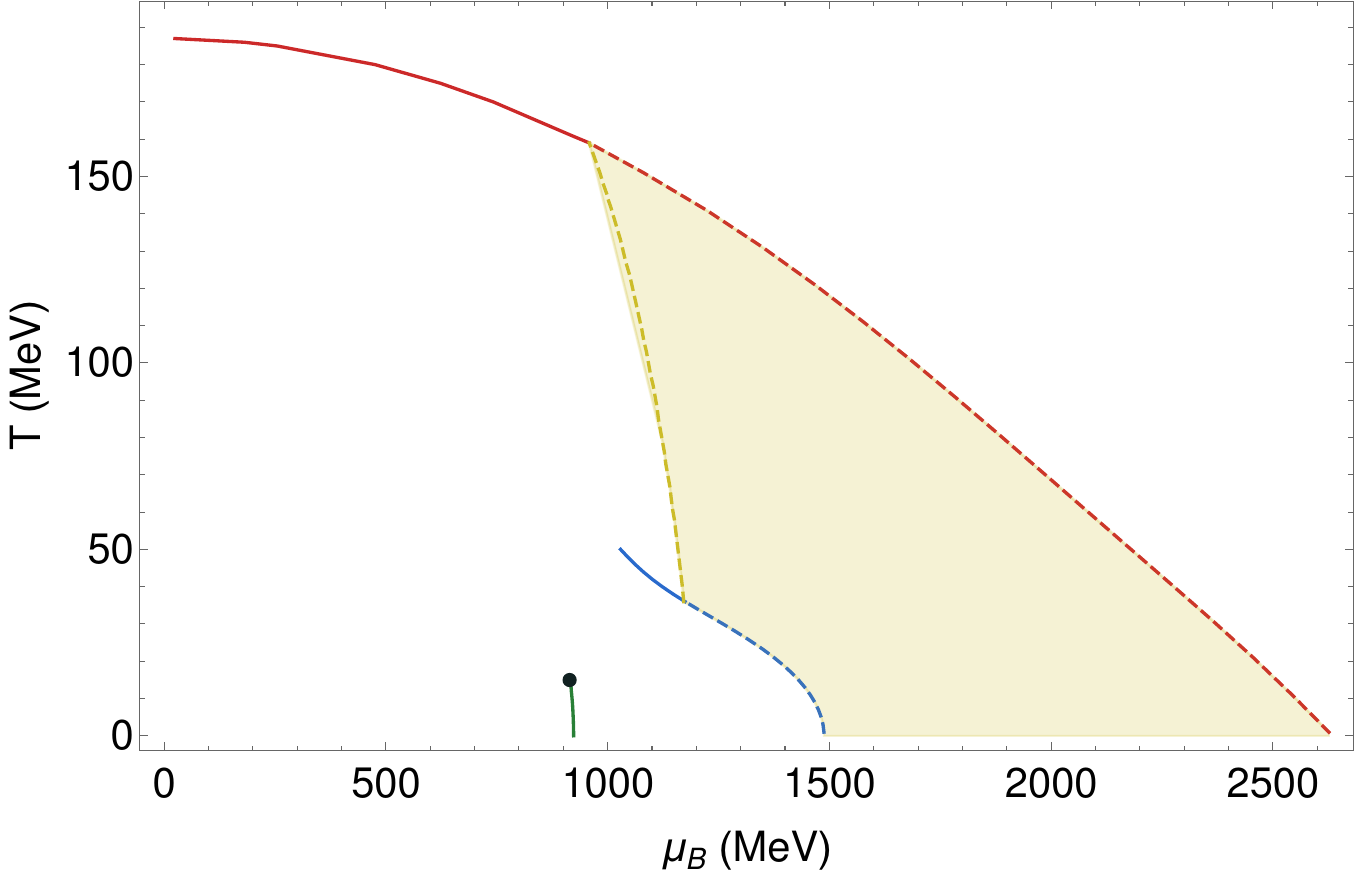}}\hfill
    \subfigure[$\zeta=1$, $g_0=27$]{\includegraphics[width=0.45\textwidth]{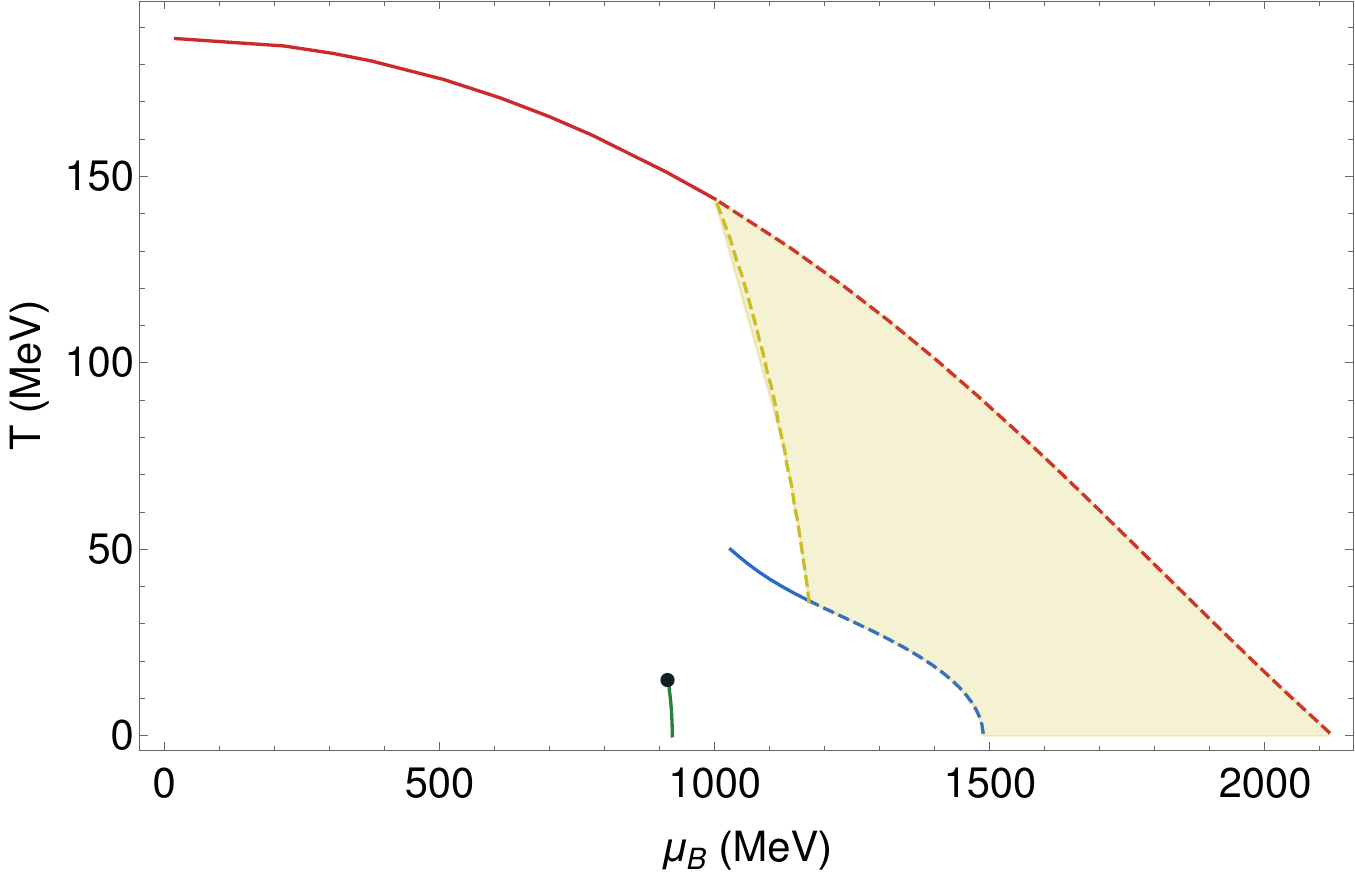}}
    \caption{Phase diagram in $\m-T$ plane for different values of $g_0$ and $\z$}
    \label{fig:2flpd}
\end{figure}

Carrying out this exercise for various temperatures gives us the phase diagram shown in figure \ref{fig:2flpd}.
In this figure, we have also depicted the van der Waals critical point of nuclear matter (obtained from the equation of state discussed in section \ref{vdw}). The red line represents a transition from the NJL description to the charged black hole phase. The blue line represents the transition from the van der Waals liquid to the quarkyonic liquid. Since the vdW description is valid only upto $T\sim50$ MeV, we do not extend this coexistence line further. The yellow shaded area indicates the region where  $\frac{\langle \psi \psi\rangle(T,\mu)}{\langle \psi \psi\rangle(0,0)}<0.1$ (chiral condensate vanishes).

It is perhaps noteworthy that the bulk description does not possess any manifest feature representing the running coupling - yet we obtain a somewhat sensible deconfinement transition.

\section{Dilaton model}\label{dilaton}

\begin{figure}[h]
    \centering
    \subfigure[Dilaton potential]{\includegraphics[width=5cm]{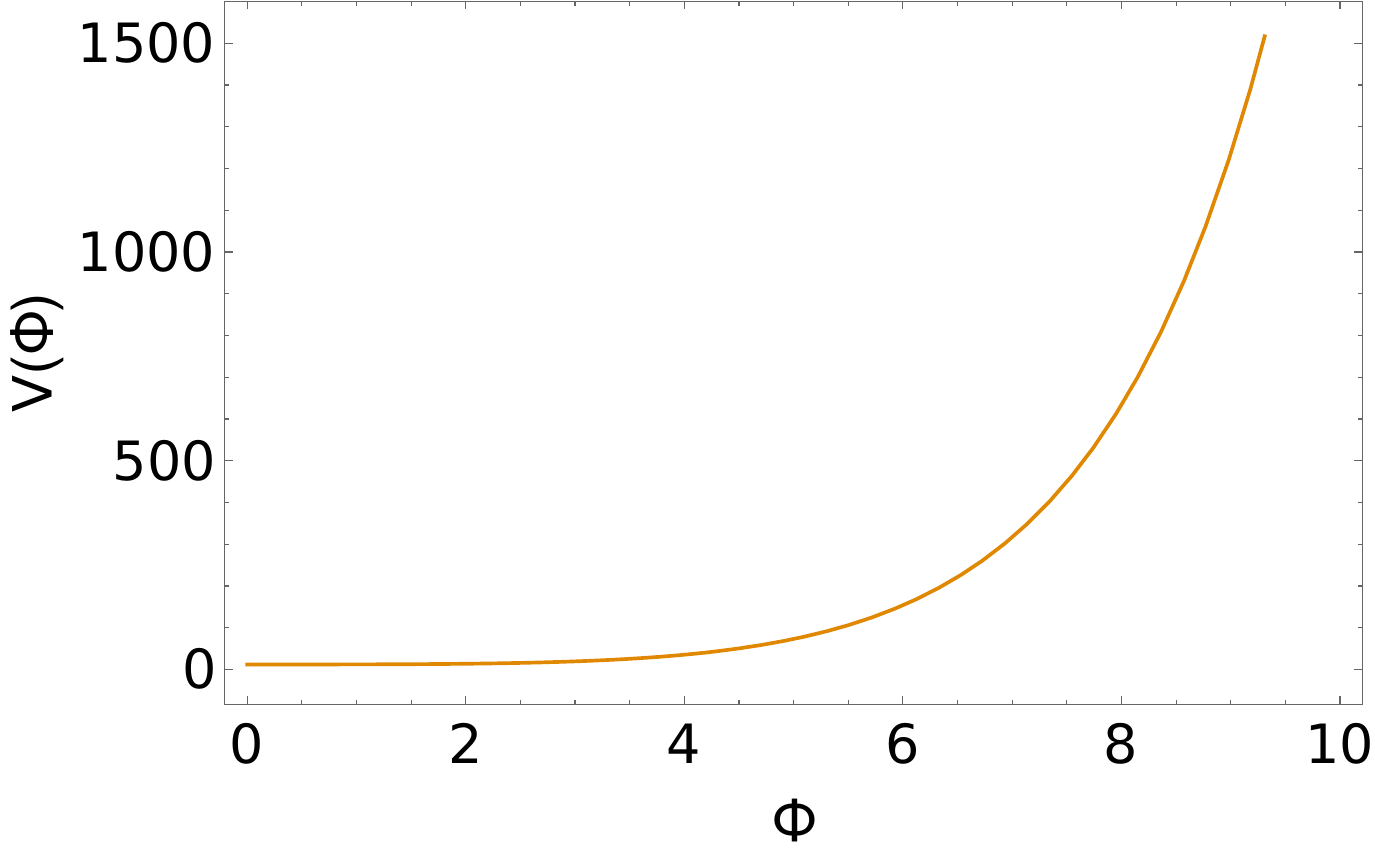}}
    \subfigure[Gauge Kinetic function]{\includegraphics[width=4.8cm]{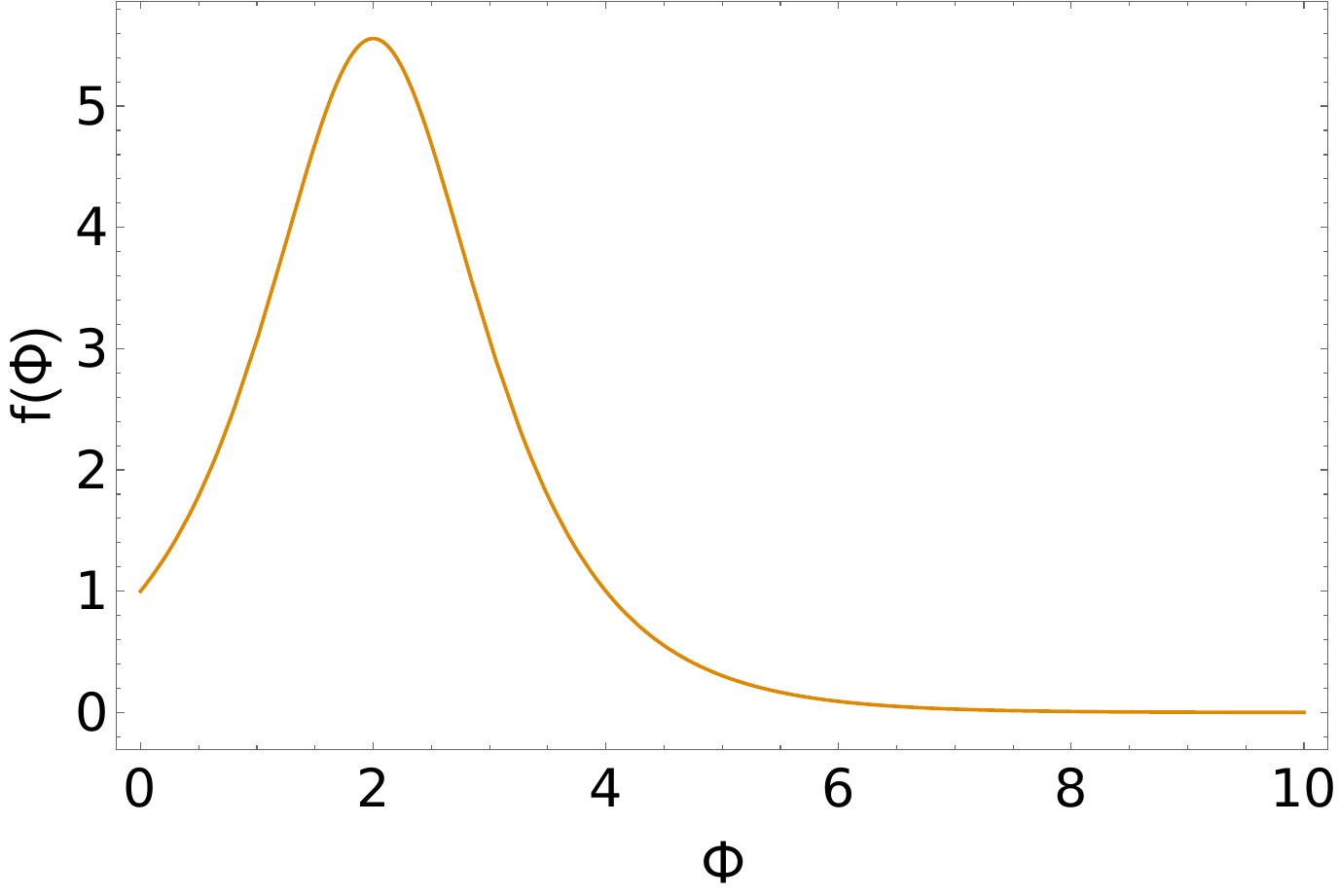}}
    \caption{Dilaton potential and Gauge Kinetic function}
    \label{fig:Pot_Dil}
\end{figure}

In applications to QCD, it is important to incorporate some version of the running coupling. 
In AdS/QCD this is done by including a bulk  dilaton field. There are several studies which incorporate gauge couplings tailored to reproduce the QCD critical point (in the $\m_B-T$ plane). In these systems as well, low density confined phases with nonzero baryon density have not been identified so far. If the proposal of the charged AdS geometries is to hold weight, we must expect to see similar confined phases at low densities in these systems too. In this section, we will consider the possibility that such a phase will occur in the model of \cite{DeWolfe:2010he}. The Action of the system is:
\be\label{actionSW}
S=\int d^5x \sqrt{g}\left[\frac{1}{2\k^2}\left(R-\frac{1}{2}(\del \Phi)^2 -V(\Phi)\right) 
- \frac{f(\Phi)}{g_5^2}\frac{F^2}{4 }\right]    
\ee
where $V(\Phi)$ and $f(\Phi)$ are the dilaton potential and the gauge kinetic function. We choose the same potential and kinetic function as given in \cite{DeWolfe:2010he}.
\be 
V(\Phi)=\frac{b \Phi ^2-12 \cosh (\g \Phi )}{L^2} \; ;\qquad f(\Phi)=\frac{\sech\left(\frac{6 (\Phi -2)}{5}\right)}{\sech\left(\frac{12}{5}\right)}
\ee 
where $\g=0.606$ and $b=2.057$. We plot the dilation potential and the gauge kinetic function in figure \ref{fig:Pot_Dil} which shows that the kinetic function vanishes for a large dilation field. 
The dilaton potential $V$ has a single minimum at $\Phi=0$ near which 
$V(\Phi)=-12+(b-6\g^2)\Phi^2$. From this, we can determine the scaling dimensions
\be
\D_\pm=\frac{d}{2}\pm \sqrt{\frac{d^2}{4}+6\g^2-b}
\ee
of the operator dual to the bulk dilaton. For the parameter values mentioned, $\D_+=3.96$ shows that we have a marginally relevant operator. We use the same bulk field ansatz as in section \ref{HW}, and for the  dilaton field, we take $\Phi=\Phi(z)$, which will give us the following equations of motion:

\begin{flalign}
&g'(z)-g(z) \left(\frac{h'(z)}{2 h(z)}+\frac{4}{z}\right)-z L^2 \frac{V(\Phi (z))}{3}  -\zeta  z f(\Phi (z)) \left(\frac{h(z) \phi '(z)^2}{2 L^2}\right)=0\\
&\frac{h'(z)}{h(z)}-\frac{4}{z}-\frac{z\Phi '(z)^2 }{3} =0\label{SWheq}\\
&\phi ''(z)-\left(\frac{3}{z}-\frac{h'(z)}{2 h(z)}+\frac{\Phi '(z) \dot{f}(\Phi (z))}{f(\Phi (z))}\right) \phi '(z)=0\\
&\Phi ''(z)+\Phi '(z) \left(\frac{g'(z)}{g(z)}-\left(\frac{h'(z)}{2 h(z)}+\frac{3}{z}\right)\right)-\frac{L^2 \dot{V}(\Phi (z))}{g(z)}+\frac{3 \zeta  \dot{f}(\Phi (z)) h(z) \phi '(z)^2}{2 L^2 g(z)}=0
\end{flalign}
where $\dot{}=\frac{\del}{\del \Phi}$.\\
The class of solutions that we will consider will involve the non-normalizable mode of the dilaton $\Phi_-$(which drives the breaking of scale invariance) being set to a common constant (in our case $\Phi_-=2 \times10^{-4}$). The non-normalizable mode of the scalar potential $\phi_{UV}=\m$. We can still use the scaling of \eqref{SWheq} to set $h(z_{UV})=z_{UV}^4.$ The remaining three boundary conditions are fixed by the regularity condition at the horizon. However, in the horizonless situation we provide these initial conditions at IR cutoff $z_0$. The solutions of these equations are then determined numerically and substituted in the action to evaluate the pressure (by adding suitable counterterms, see Appendix \ref{dil_CT}).

The charged black hole solutions with a nontrivial dilaton profile are determined by specifying the remaining boundary conditions at the horizon. We fix the metric function $g(z_H)=0$ by assumption, the scalar potential $\f(z_H)=0$ due to the time circle shrinking to zero size at the horizon, and regularity of the equation of motion at the horizon is ensured by setting the derivative of the dilaton field $\Phi'(z_H)=\frac{1}{2 L^2 g'(z_H)}\left( \dot{V}(\Phi (z_H)) L^4 -3 \zeta  \dot{f}(\Phi (z_H)) h(z_H) \phi '(z_H)^2 \right)$.  We can vary the temperature via the only free parameter, i.e., the horizon $z_H$. For a given chemical potential and temperature value, the black hole solution is completely determined by these boundary conditions.

We note that for extremal black holes, the boundary condition for the scalar potential changes to
\[\f'(z_H)^2=-\frac{2L^4 V(\Phi(z_H))}{3\z h(z_H)f(\Phi(z_H))}; \qquad \f(z_H)=0\] 
therefore, we cannot use the $\f$ boundary conditions to fix the chemical potential at the boundary. However, we can vary the horizon $z_H$ to set $\f(z_{UV})=\m.$ Similarly, we obtain a relation between the dilaton potential and the gauge kinetic function enforced by the zero temperature (or $g'(z_H)=0$) condition,
\[\frac{\del }{\del\Phi}\left(log(V (\Phi) f(\Phi))\right)\Big|_{z=z_H}=0\]
this fixes dilaton field $\Phi(z_H)=2.182$. The $\Phi'(z_H)$ is now the free parameter, which we will fix by the non-normalizable mode of the dilaton field $\Phi_-=2 \times10^{-4}$ at the UV boundary.

To obtain the horizonless solutions of charged AdS (CAdS) type with a nontrivial dilaton field, we first note that the gauge kinetic function acts as a potential for the dilaton. Therefore, the natural boundary condition for the dilaton is the one that minimizes the on-shell action at $z_0$. As discussed in the previous sections, we set $g(z_0)=6 z_0^2$. For the scalar potential, we first take $\f(z)=\m$ and find that the constant scalar potential solutions always are the lowest pressure phase. Therefore, we will not discuss these geometries further. For the finite density confined geometries, we again take the simplest boundary condition $\f(z_0)=0$. We specify all boundary conditions at the IR cutoff $z_0$ and use the shooting method to match the desired value at the UV boundary. 
\begin{figure}[h]
    \centering
    \includegraphics[width=0.24\textwidth]{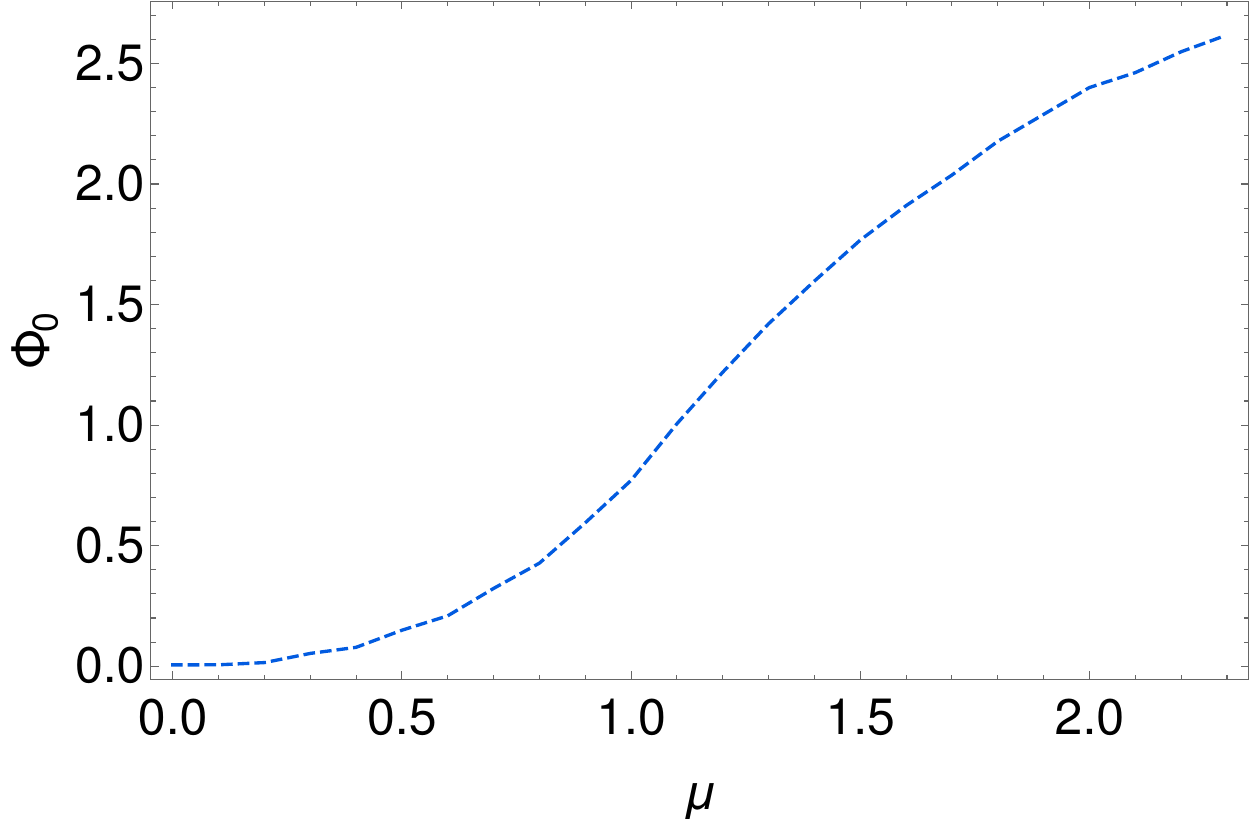}
     \includegraphics[width=0.24\textwidth]{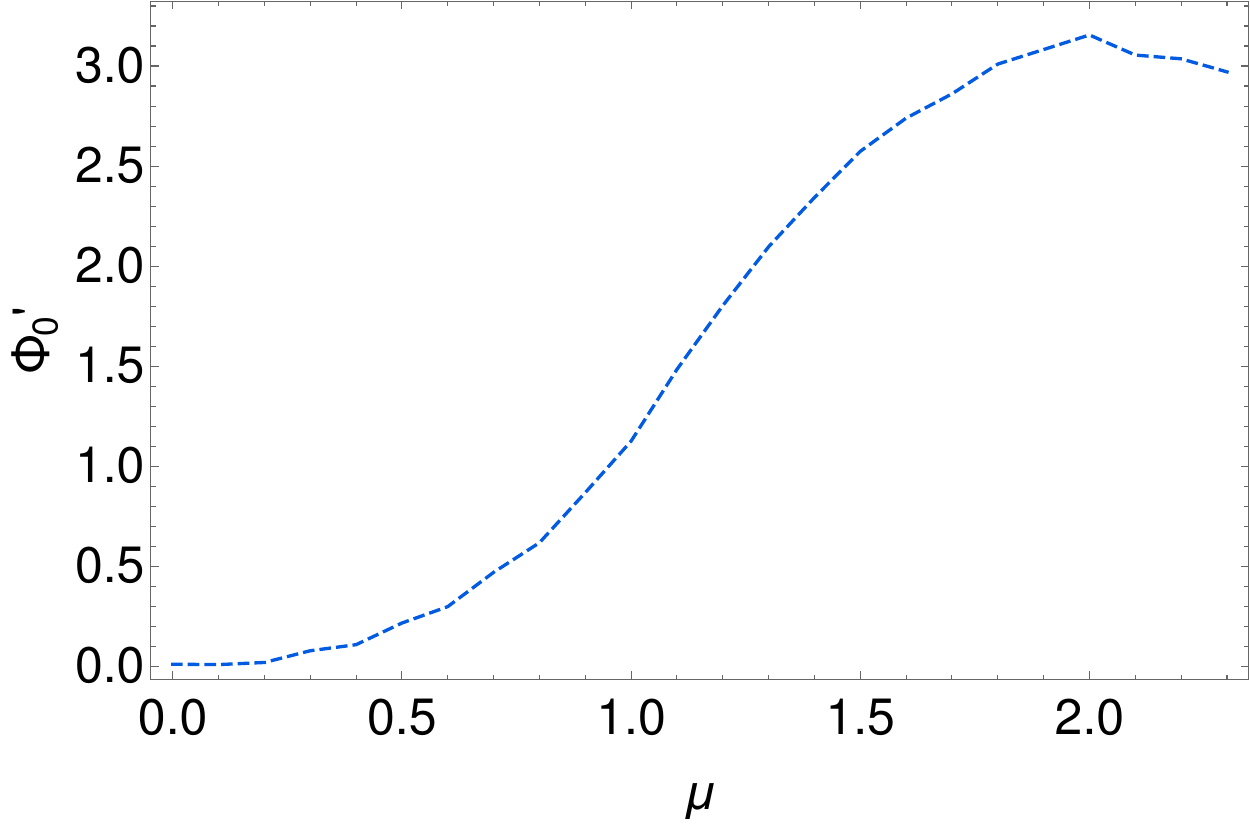}
    \includegraphics[width=0.25\textwidth]{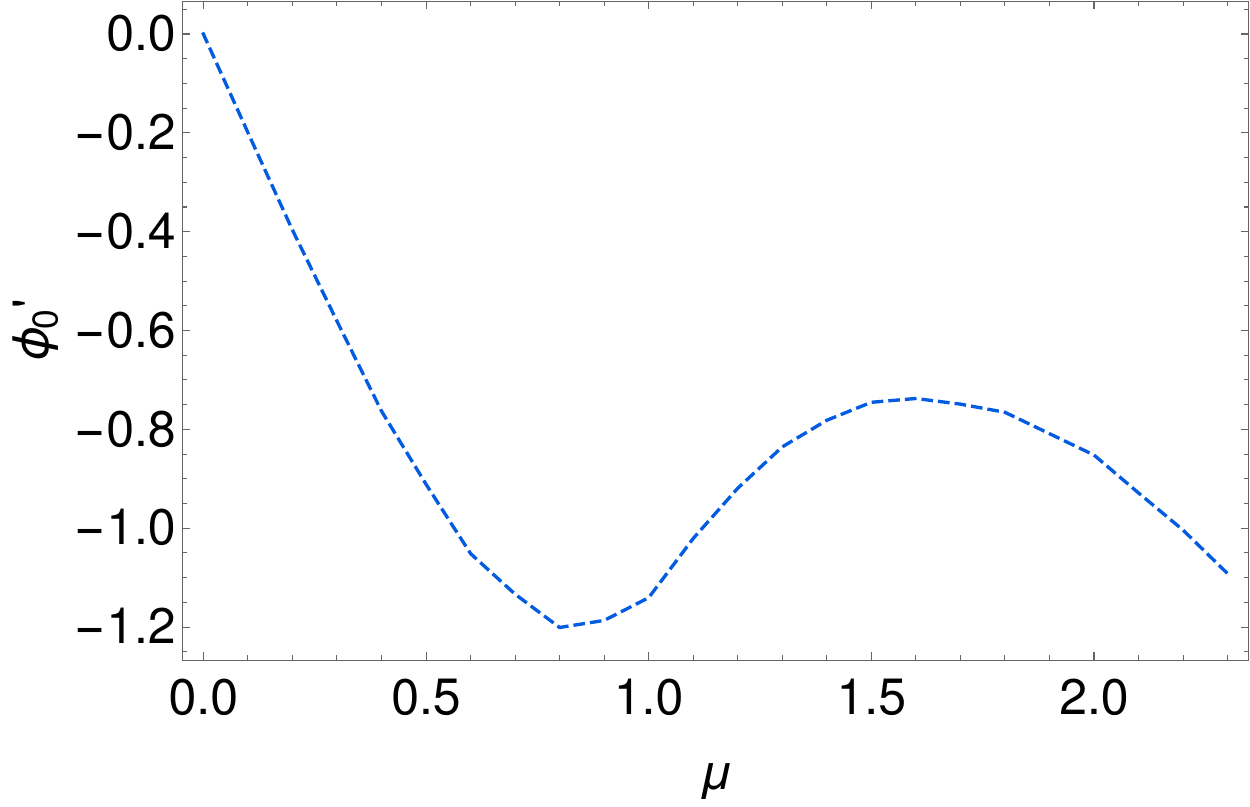}
    \includegraphics[width=0.24\textwidth]{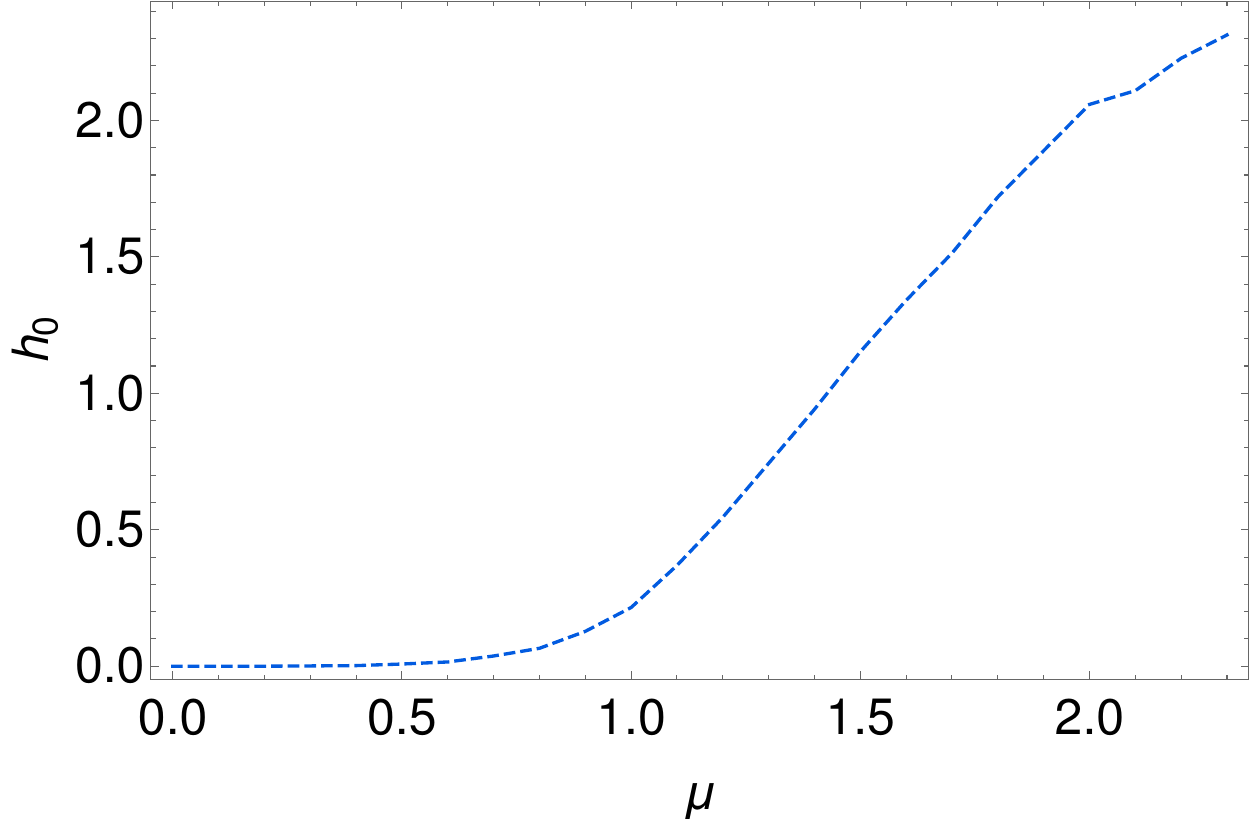}
    \caption{The initial conditions at the IR cutoff $z_0$ (that have minimum energy) as a function of dimensionless chemical potential $\m$}
    \label{fig:IC}
\end{figure}

Figure \ref{fig:IC} shows the initial conditions for the dilaton field $\Phi(z_0)=\Phi_0$, the derivative of the dilaton field $\Phi'(z_0)=\Phi_0'$, the electric field $\f'(z_0)=\f'_0$, the metric function $h(z_0)=h_0$ at the IR cutoff $z_0$ for the minimum on-shell action (energy) solutions as a function of the dimensionless chemical potential. We observe that for large chemical potential, after the maximum in the initial condition of the electric field and the derivative of the dilaton field, the solution converges slowly. Consequently, the phase transition line does not exhibit smooth behavior for large chemical potentials, as shown in figure \ref{fig:SW_PD} below.

\begin{figure}[h]
    \centering
    \subfigure[$\m=0.1$]{\includegraphics[width=0.4\linewidth]{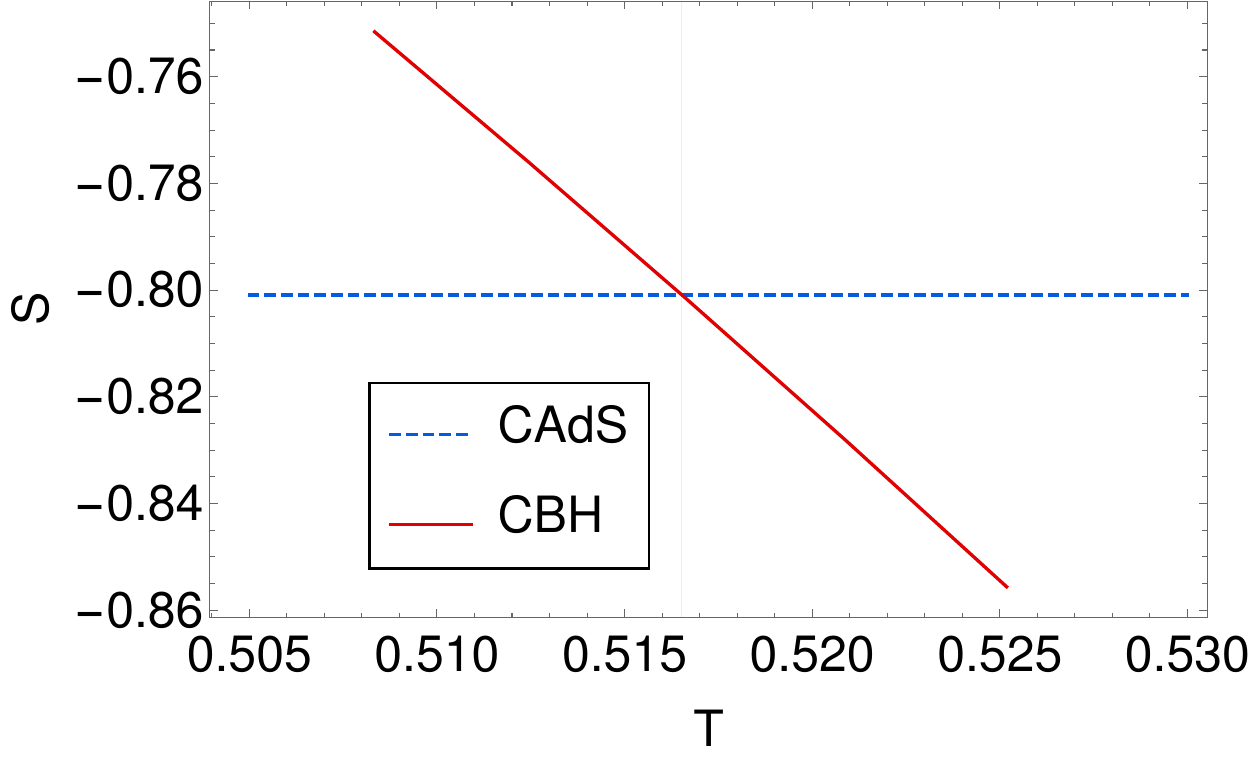}}
    \subfigure[$\m=1$]{\includegraphics[width=0.4\linewidth]{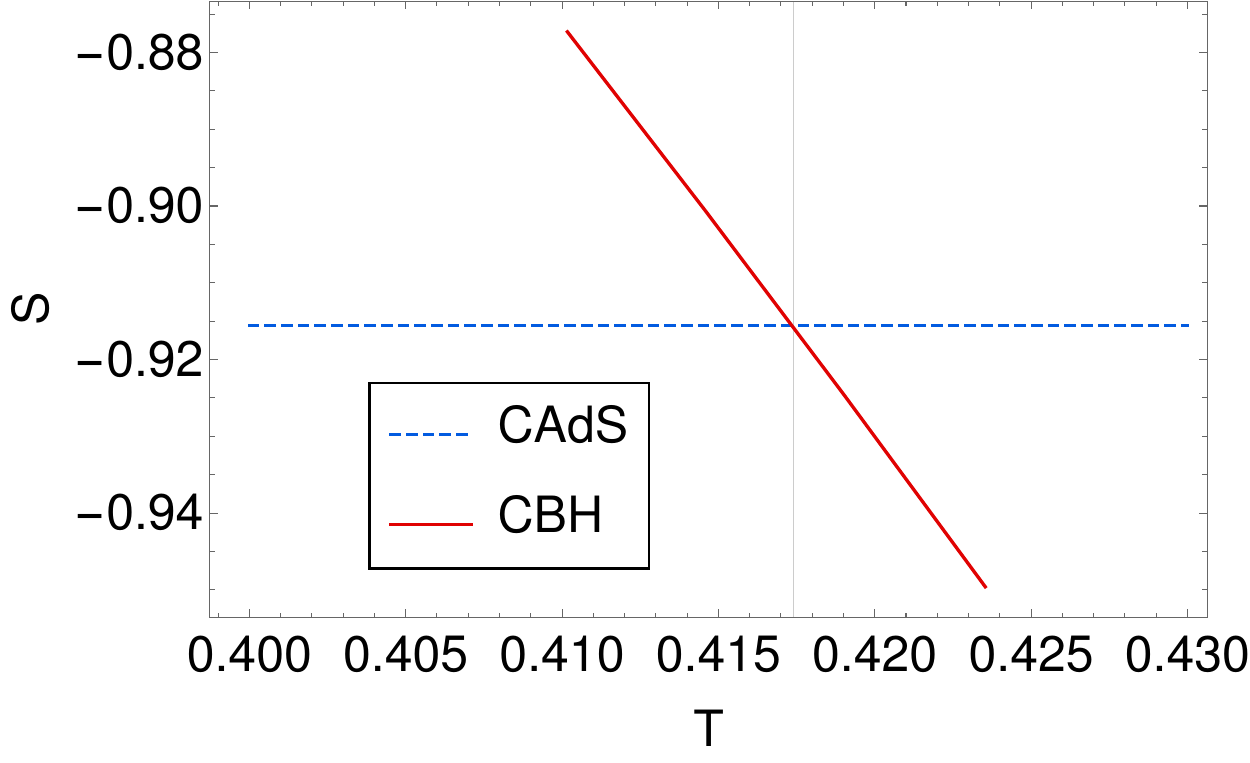}}
    \caption{The on-shell action for CAdS and CBH (with dilaton) as a function of $T$ for different values of the chemical potential.}
    \label{fig:SW_PT_T}
\end{figure}

We now compare the CAdS solution with the CBH solutions for fixed dimensionless chemical potential. In figure \ref{fig:SW_PT_T}, we show the on-shell action for the CAdS (represented by the blue dashed line) and the CBH (represented by the solid red line). The on-shell action is independent of temperature for the CAdS geometries. We again find that there is a confinement/deconfinement phase transition with a nontrivial dilaton profile in the hardwall model.

\begin{figure}[h]
    \centering
    \includegraphics[width=0.45\textwidth]{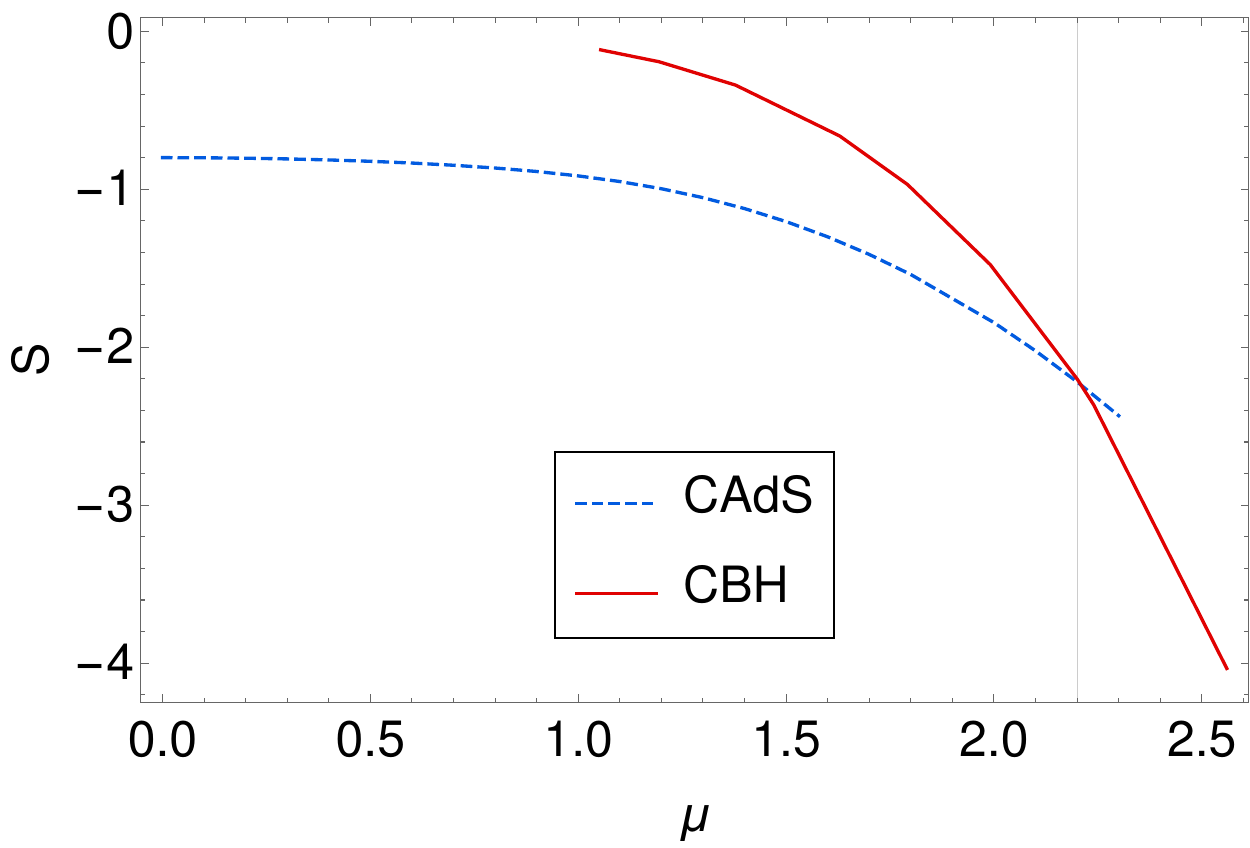}
    \caption{On-shell action for CAdS and CBH (with dilaton) at $T=0$}
    \label{fig:SW_ZERO}
\end{figure}

In figure \ref{fig:SW_ZERO}, we plot the on-shell action as a function of the dimensionless chemical potential $\m$. Again the dashed blue curve represents the CAdS geometry with dilaton. However, the red curve shows the on-shell action for extremal black holes ($T=0$) as a function of the dimensionless chemical potential. The confinement/deconfinement phase transition occurs at $\m\simeq 2.2$.

\begin{figure}[h]
    \centering
    \includegraphics[width=0.45\textwidth]{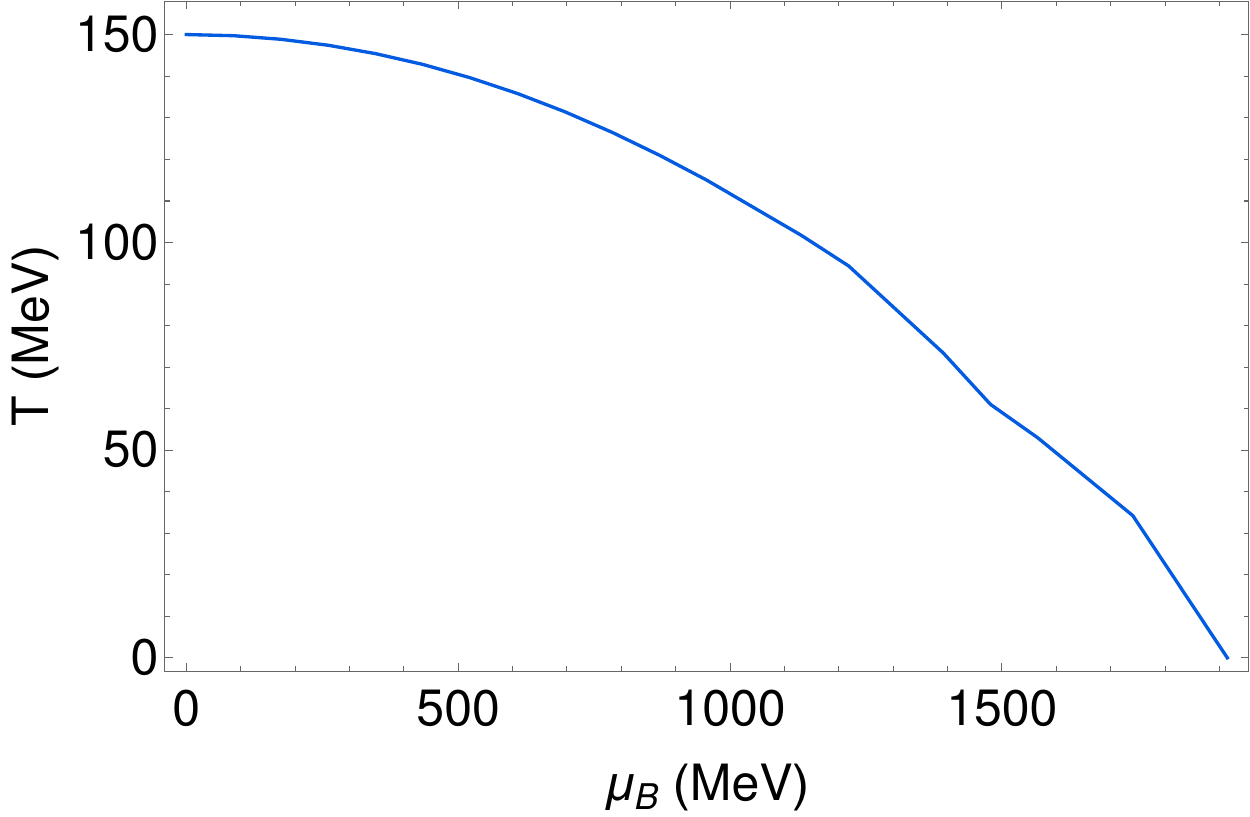}
    \caption{Phase diagram in $\m-T$ plane with a nontrivial dilaton field.}
    \label{fig:SW_PD}
\end{figure}

Thus, we conclude that there is a deconfinement phase transition between the black hole type solutions and the solutions without a horizon. Repeating this for a few values of the temperature gives us the figure shown in figure \ref{fig:SW_PD} where the phase coexistence line is drawn in blue. The phase diagram is represented in the physical units using the cutoff $z_0$ as discussed in section \ref{HW}.

\section{Discussion}
In this work, we study asymptotically AdS background geometries with nonzero electric fields (termed CAdS) and a hardwall cutoff as the gravitational dual of a phase with confined baryonic matter (at finite density and temperature). We have shown that upon including these geometries, we obtain satisfactory phase diagrams for QCD applications.

Another key contribution of this work is the use of phenomenologically motivated boundary conditions that allowed us to match the equations of state with low density effective field theory results. This enabled us to create a comprehensive phase diagram, including phase transitions such as the van der Waals liquid-gas transition for nuclear matter and the quarkyonic-baryonic liquid transition (chiral symmetry restoration) indicated by chiral condensate vanishing.

The hardwall approach throws up intriguing inequalities like \eqref{muCRIT_QCD} (similar inequalities were also obtained in the 10-D model in \cite{Singh:2022obu}) that could be insightful for QCD applications. This is primarily due to the fact that these inequalities illuminate the dependence of the glueball mass and the vector meson mass on the chemical potential, above which the deconfined phase becomes favored at zero temperature. 
We also note that the phase diagram depends sensitively on the parameter $\q$. There is a small range of $\q$ for which the phase diagram matches the expected QCD results. From a 10-D perspective, this can mean that there are certain types of compactification and dilaton coupling that will provide a sensible dual description of the boundary theory. Within this parameter range, in agreement with \cite{Costa:2019bua}, the phase diagram suggests that at a fixed value of temperature, as we increase the chemical potential, the chiral restoration appears first, signaled by the vanishing of the chiral condensate. Subsequently, the quark mass becomes equivalent to the current quark mass, followed by the melting of the quarks into the deconfined phase as the chemical potential is increased further indicated by the entropy density $\propto \mathcal{O}(N_c^2)$. 

In QCD, the running coupling is known to play a crucial role in determining the low energy phases. Therefore, in the second part, we studied a particular dilaton model \cite{DeWolfe:2010he} within the hardwall setting, where again we have shown that horizonless geometries contribute a low density confined region to the phase diagram. We have argued that, at finite density, the natural boundary condition for the dilaton field in the IR region is determined dynamically by minimizing the on-shell action. The phase diagram in the $\m-T$ plane with a nontrivial dilaton profile is shown in figure \ref{fig:SW_PD}.  The original model was constructed to reproduce the QCD critical point. This occurs because there are multiple bulk solutions for a given temperature and chemical potential. In our work, to exhibit proof of principle, it was sufficient to find one class of bulk backgrounds. 
Clearly a future effort should attempt to identify the QCD critical point as well. A more realistic QCD-like model can include finite isospin and also the bulk fields dual to chiral condensates and possibly baryon condensates as well. Such a study will be of great interest since one can make closer contact with QCD phenomenology.

These geometries have growing electric fields in the IR regions. Introducing an IR cutoff will excise these regions, and, therefore, allow us to work in the region of validity of the action. If the cutoff is chosen so that $\alpha' F <<1 $, these geometries are sensible as string solutions in that higher derivative terms will not modify these drastically. The growing electric field makes these models similar to the softwall models in which the gravitational curvature diverges in the IR. 

However, we also emphasize that this result should be regarded as a proof of principle that there exists a backreacted finite density confined geometry (with running of the boundary coupling). We expect that there exists an IR complete geometry that will approach the same metric in between the UV and the IR cutoffs, and the phase diagram and equations of state obtained with this complete geometry will remain qualitatively equivalent to those presented in this work.  In fact, in a very interesting recent work,\cite{Faedo:2023nuc} has constructed a holographic dual corresponding to this phase albeit in 2+1 boundary dimensions.

A possible source of concern is that the CAdS geometries are obtained by truncating over charged black holes in AdS. Such black holes are known to possess thermodynamic instabilities. It is important to ensure that due to the IR boundary conditions such instabilities are removed. 

The advantages of the hardwall model are somewhat cancelled by the arbitrariness of the IR boundary conditions. It will be interesting to systematically explore the IR boundary conditions and their role in the AdS/CFT correspondence. There are two guiding principles that can orient such studies - the first is along the lines of Holographic renormalization group which can be regarded as the evolution of the boundary conditions from $\e$ to $z_0.$ 
The second guide is diffeomorphism invariance (and other gauge symmetries of the bulk) - which in field theory becomes Ward identities involving the stress tensor. The IR boundary conditions must be expressible in terms of diffeo-invariant quantities so that they can be given precise boundary meaning. Systematizing these so that the boundary Lorentz invariance is intact can allow us to introduce specific sources to capture known physics and perhaps predict new phenomena.

The major use of such backreacted finite density geometries is to obtain the equation of state in extreme conditions like inside compact stars. In these situations exotic phases condensates, superconductors, and the transport coefficients are hard to compute from the boundary field theory. However, using these backgrounds, we can study those systems by adding relevant degrees of freedom in the bulk action via appropriate fields \cite{McGreevy:2009xe,Blake:2022uyo,Hartnoll:2008kx}. For instant, neutron pairing inside the compact stars \cite{Pethick:2015jma} can be studied by analysing the CAdS type geometry with a complex scalar field \cite{Hartnoll:2008kx} to model the neutron condensate.

Finally, it will be important to compare the hardwall models with complete 10-D string theoretic models such as the D3/D7, Klebanov-Strassler or the Witten-Sakai-Sugimoto models to formulate better strategies toward a QCD phase diagram. In these cases, 
we can have flavor branes with world-volume electric fields embedded in black hole or AdS backgrounds.  Presumably the CAdS backgrounds discussed here correspond to the situations with no horizon within the flavor brane worldvolumes. In the former situation, the presence of a horizon in the induced worldvolume metric can signal a ``deconfinement" of the flavor degrees of freedom distinct from that of the gluons. Even in 5-d, it is important to compare with the highly studied IHQCD and VQCD models \cite{Jarvinen:2021jbd} to extract universal features of the holographic approach and confront those features with QCD investigations proper (such as those based on Lattice QCD or effective field theories). The NJL and van der Waals degrees of freedom behind the IR cutoff of this work are likely to be modelled in these setups in terms of an interacting gas of baryonic degrees. In the 10-D perspective these baryons could be represented by wrapped D-branes or bound-states of fundamental strings. It will be highly interesting to explore these situations and assemble a coherent description of these phases of QCD.

\appendix

\section{Holographic renormalization}
\label{holoren}
The action \eqref{action1} yields UV divergences when evaluated on-shell from the gravity components. To address this, we have implemented the holographic renormalization scheme \cite{deHaro:2000vlm}, which adds additional terms that we discuss below.

 \begin{align}
    S_{ren}=S_g+S_M+S_{GH}+S_{ct1}\label{S_ren}\\
    S_{GH}=-\frac{2}{2\k^2}\int d^{4}x\sqrt{\g}\theta\\
    S_{ct1}=\frac{2}{2\k^2}\int d^{4}x\sqrt{\g}\frac{3}{L}\\
\end{align}

\begin{itemize}
    \item The Gibbons-Hawking term is added for well-definedness of the variational principle. We do not add a GH term at the IR boundary because this is not a real boundary.  Here $\g$ is the induced metric on the UV surface $z=\e$ and $\theta$ is the trace of the extrinsic curvature associated with this surface.
    \item The counter term $S_{ct1}$ cancels divergences from the gravity part $S_g$ as well as from Gibbons-Hawking term $S_{GH}$. 
    \item In the matter part $S_M$, the field strength $F^2$ never diverges because of the asymptotic nature of the gauge field in the UV.     
\end{itemize}

For the given field ansatz we can evaluate terms in the action as:
\be
\theta=\frac{\sqrt{g(z)}}{L}\left( \frac{h'(z)}{2h(z)}+\frac{3}{z}-\frac{g'(z)}{2g(z)}\right); \quad
\sqrt{\g}=\frac{ L^4}{z^3}\sqrt{\frac{g(z)}{h(z)}}; \quad
\sqrt{g}=\frac{L^5}{z^3}\sqrt{\frac{1}{h(z)}}
\ee

\be
S_{g}+S_{M}=-\frac{\b V L^3}{2\k^2}\left[\frac{2}{z ^4}\frac{g(z)}{\sqrt{h(z)}}\right]_{z=\e}^{z=z_{IR}}
\ee
\be
S_{GH}=-\frac{L^3}{2\k^2}\int d^{4}x \frac{g(z)}{z^3\sqrt{h(z)}}\left( \frac{h'(z)}{2h(z)}+\frac{3}{z}-\frac{g'(z)}{2 g(z)}\right)
\ee
\be
S_{ct1}=\frac{3 L^3}{\k^2}\int d^{4}x \frac{1 }{z^3}\sqrt{\frac{g(z)}{h(z)}}
\ee

The on-shell Euclidean action for \textbf{CBH} is
\be
S_{CBH}=-\frac{\b V L^3}{2\k^2}\left[-\frac{2}{\e^4}+2\left(\frac{1}{z_H^4}+\frac{\z}{L^2}\frac{\m^2}{z_H^2}\right) \right]
\ee
where the IR terms are assumed to vanish for the black hole. 

The Gibbons-Hawking and Counter term are:
\be
S_{GH}=-\frac{\b V L^3}{2\k^2}\left[\frac{8}{\e^4}-4 \left(\frac{1}{z_H^4}+\frac{\z}{L^2}\frac{\m^2}{z_H^2}\right)\right]; \qquad
S_{ct1}=-\frac{\b V L^3}{2\k^2}\left[-\frac{6}{\e^4}+3 \left(\frac{1}{z_H^4}+\frac{\z}{L^2}\frac{\m^2}{z_H^2}\right)\right]
\ee

The renormalized on-shell action is
\be
S_{CBH}^{ren}=S_{CBH}+S_{GH}+S_{ct1}
=-\frac{\b V L^3}{2\k^2}\left[\frac{1}{z_H^4}+\frac{\z}{L^2}\frac{\m^2}{z_H^2}\right]
\ee

Similary one can compute the on-shell Euclidean action for \textbf{CAdS}:
\be
S_{CAdS}^{ren}=-\frac{\b V L^3}{2\k^2}\left[c\red{+\frac{2g(z_0)}{z_0 ^6}}\right]
\ee
and the IR terms are shown in red.
For the CAdS $g(z_0)= z_0^2 (1-c z_0 ^4 +\frac{\z}{L^2} Q^2 z_0 ^6)$, so 
\be
S_{CAdS}^{ren}=-\frac{\b V L^3}{2\k^2}\left[\frac{1}{z_0^4}+\frac{\z}{L^2} Q^2 z_0^2+\frac{g(z_0)}{z_0 ^6}\right]
\ee

\subsection{The Dilaton Counter Term}\label{dil_CT}

In addition to the Einstein-Hilbert component of the action, the dilaton also sources UV divergences. We again implement the holographic renomalisation scheme described in \cite{deHaro:2000vlm}. A further addition of a counter term is introduced apart from what is described in the previous section. 
\be 
S_{ct2}=\frac{1}{2\k^2}\int d^{4}x\sqrt{\g} \Phi^2 \frac{\left(d-\delta _+\right)}{2L}
\ee

\bibliography{p1}
\end{document}